\documentstyle[12pt,epsf,graphics]{article}
\begin{document}
\title{Shortcuts in a Nonlinear Dynamical Braneworld in Six Dimensions} 
\author{Bertha Cuadros-Melgar \footnote{e-mail: bertha@fma.if.usp.br}\\ \\
Instituto de F\'\i sica, Universidade de S\~ao Paulo \\
  C.P.66.318, CEP 05315-970, S\~ao Paulo, Brazil}
\date{}
\maketitle

\begin{abstract}
We consider a dynamical brane world in a six-dimensional spacetime
containing a singularity. Using the Israel conditions we study the
motion of a 4-brane embedded in this setup. We analyse the brane
behaviour when its position is perturbed about a fixed point and solve
the full nonlinear  
dynamics in the several possible scenarios. We also investigate
the possible gravitational shortcuts and
calculate the delay between graviton and photon signals
and the ratio of the corresponding subtended horizons.
\end{abstract}


\newpage

\section{Introduction} 

It has recently been argued that there might exist some extra spatial
dimensions, not in the traditional Kaluza-Klein (KK) scheme \cite{KK} where
extra dimensions are compactified on a radius of the order of Planck
scale, but in a scenario where they could be large. Certainly, the extra
dimensions idea is not completely new but it is also related to string
theories. Among the existing models the ten-dimensional $E_8 \times
E_8$ heterotic string theory seems to be the most acceptable candidate
to describe our world. This theory is related to an 11-dimensional
spacetime, where the 11th dimension is compactified via a $Z_2$
orbifold symmetry. In this setup the standard model particles are
confined to a four-dimensional spacetime, while the gravitons
propagate in full spacetime. 

Large extra dimensions were also recalled to solve the hierarchy
problem \cite{arkani}. However, it was the work of Randall and Sundrum
\cite{rs1,rs2} which suscitated a renewed interest on these grounds. In
this model our world is identified with a domain wall in
five-dimensional anti de Sitter (AdS) spacetime. In their first paper,
Randall and Sundrum proposed a mechanism to solve the hierarchy
problem by a small extra dimension, while in their second paper, the
braneworld with a positive tension was investigated. 

In this paper we consider a six-dimensional model where the bulk
contains a singularity and is bounded by a four-dimensional
brane containing matter. This brane is a thick brane with the usual three
infinite spatial directions and an additional coordinate compactified in a
small radius that can be of Planck size like in the KK models
\cite{KK} or even of submilimetric size as in the Arkani-Hamed,
Dimopoulos and Dvali (ADD) models \cite{arkani}, which makes it unavailable
to the brane observers. In this way, this ``effective'' picture
corresponds to our universe. 
The static case has been already studied in \cite{accm} where 
AdS-Schwarzschild and AdS-Reissner-Nordstr\"om embeddings were
considered. The case of domain walls moving in bulks containing
non-charged singularities has also been studied in \cite{kraus,chr}. Brane
models in AdS space with Schwarzschild singularities 
have been used to understand the AdS/CFT correspondence and look like
promising theoretical models \cite{aw}.

The motivation for a six-dimensional model comes from the fact that
spacetimes with more than one extra dimension can allow for solutions
with more appealing features, particularly in spacetimes where the
curvature of the internal space is non-zero. More extra dimensions
also relax the fine-tunnings of the fundamental parameters. On the
other hand, the ADD model \cite{arkani}, which 
relates the Planck mass to the fundamental mass in $(4+n)$ dimensions
through the size of the transverse dimensions, foresees the existence
of $n\geq 2$ extra dimensions. The possibility $n=1$ remains
excluded from experimental bounds since there are no observed
violations of the Newton's law at distances of the order of the solar
system.

According to the model, photons are confined to the brane and
gravitons can propagate in all five spatial directions, then there is
a possibility that gravitational fields while propagating out of the
brane speed up due to the warped bulk geometry reaching farther
distances in smaller time than light propagating inside the brane, a
scenario that for a resident of the brane (as ourselves) implies
shortcuts \cite{acmfw,accm,csaki,5sc,cf}.

The existence of shortcuts implies that causality is violated from the
point of view of a four-dimensional observer; however, it is
perfectly defined through $N$-dimensional geodesics for an observer in
the bulk. Thus, two points apparently causally disconnected in four 
dimensions could be causally connected by gravitational
shortcuts. This fact accounts for a possible solution of the
horizon problem. As it was pointed out in \cite{ac}, if high redshifts
were available, shortcuts appearing before nucleosynthesis are serious
mediators of homogenization of the matter on the brane and then they
could provide
an explanation alternative to inflation to this problem. Other
alternative approaches have also considered a varying speed of light
\cite{mag}. Moreover, if
inflation took part on the brane, the causal structure is definitely
changed by shortcuts possibly leading to a non-usual period of causal
evolution of scales, what could be responsible for distinct
predictions in the cosmic microwave background structure for
inflationary models.

The paper is organized as follows. In section 2 we describe the
six-dimensional model considered along this work, derive the
equation 
system governing the brane motion from the point of view of a bulk
observer and write the geodesic equation corresponding to the shortest
graviton path in the bulk as well as the expression for the
delay between graviton and photon flight times. Section 3 is devoted
to studying the brane behaviour when its position is perturbed about a
fixed point. In section 4 we provide the solutions of geodesic
equation and brane equation of motion in the different scenarios
resulting from the combination of the parameters appearing in the
model and show the existence of several shortcuts. We also find the
time delay and the ratio between the horizons subtended by
gravitons and photons. Finally, we discuss our results in section 5.

\section{The Brane Cosmological Model}

We consider a six-dimensional model described by the following metric 
\begin{equation}\label{metric}
ds^2 = -n^2(t,y,z) dt^2 + a^2(t,y,z) d\Sigma_{k} ^2 + b^2(t,y,z)
dy^2 + c^2(t,y,z) dz^2 \, ,
\end{equation}
where $d\Sigma_{k} ^2$ represents the metric of the three-dimensional
spatial sections with $k=-1,\,0,\,1$ corresponding to a hyperbolic, a
flat and an elliptic space, respectively.

The matter content on the brane is directly related to the jump of the
extrinsic curvature tensor across the brane \cite{chr,radion}. This
relation has been derived in the case of a static brane in a previous work
\cite{accm}. Here we generalize our result 
for the Israel conditions to include the case of a brane moving with
respect to the coordinate system. Its position at any bulk time $t$ will be
denoted by
\begin{equation}\label{z}
z = {\cal R}(t) \, .
\end{equation}

The extrinsic curvature tensor on the brane is given by
\begin{equation}\label{excurv}
K_{MN} = \eta^L _M \bigtriangledown _L \tilde n_N \, ,
\end{equation}
where $\tilde n^A$ is a unit vector field normal to the brane
worldsheet and
\begin{equation}\label{eta}
\eta_{MN} = g_{MN} - \tilde n_M \tilde n_N
\end{equation}
is the induced metric on the brane.

In order to compute the components of $\tilde n^A$, we use the
relations
\begin{equation}\label{rel}
g_{MN} \tilde n^M \tilde n^N =1 \quad , \quad g_{MN} \tilde n^M u^N =0
\, ,
\end{equation}
where we have introduced the unit velocity vector corresponding to the
brane, which reads
\begin{equation}\label{u}
u^A = \left\{ {{dt}\over {d\tau}}, 0,0,0,0,{{dz}\over{d\tau}} \right\}
\, .
\end{equation}

The relation between $dt$ (the bulk time) and $d\tau$ (the brane time)
can be found from the induced metric,
\begin{eqnarray}\label{induced}
ds^2_{induced} &=& - \left[ n^2(t,{\cal R}(t)) -c^2(t,{\cal R}(t)) \dot
z^2 \right] dt^2 + a^2(t,{\cal R}(t)) d\Sigma^2 _{(4)} \nonumber \\
&=& -d\tau^2 + a^2 (\tau) d\Sigma^2 _{(4)} \, ,
\end{eqnarray}
where a dot means derivative with respect to the bulk time $t$. We
obtain 
\begin{equation}\label{dtdtau}
d\tau = n(t,{\cal R}(t)) \sqrt{ 1- {{c^2(t,{\cal R}(t))}\over
{n^2(t,{\cal R}(t))} } \dot {\cal R}^2}\, dt \equiv n \gamma^{-1} dt \, .
\end{equation}
Thus, (\ref{u}) can be written as
\begin{equation}\label{unew}
u^A = {\gamma \over n} \left\{ 1,0,0,0,0,\dot{\cal R}\right\}\, ,
\end{equation}
and from (\ref{rel}) we can easily obtain
\begin{equation}\label{ntilde}
\tilde n^A = \left\{ {{c \, \dot{\cal R}}\over {n^2
\sqrt{1-{{c^2}\over{n^2}} \dot{\cal R}^2}}},0,0,0,0,{1 \over
{c\sqrt{1-{{c^2}\over{n^2}} \dot{\cal R}^2 }}} \right\} \,. 
\end{equation}

Now we can calculate the components of the extrinsic curvature tensor
by substituting (\ref{ntilde}) into (\ref{excurv}). The non-zero
components are
\begin{eqnarray}\label{curvature}
K_0 ^0 &=& {c \over n^2} \left( 1- {{c^2(t,{\cal R}(t))}\over
{n^2(t,{\cal R}(t))} } \right)^{-5/2} \times \nonumber \\
&& \times \left\{ \ddot{\cal R} + {{nn'}\over c^2} - \dot{\cal R} \left({\dot n
\over n} -2 {\dot c \over c} \right) - \dot{\cal R}^2 \left( 2
{{n'}\over n} - {{c'}\over c} \right) - \dot{\cal R}^3 {{c\dot c}\over
n^2} \right\} \, ,\\
K^6 _0 &=& \dot{\cal R} K^0 _0 \, , \quad \quad \quad K^0 _6 =
-{{c^2}\over n^2} \dot{\cal R} k^0 _0 \, ,\\
K_i ^j &=& {1 \over c} \left( 1- {{c^2(t,{\cal R}(t))}\over
{n^2(t,{\cal R}(t))} } \right)^{-1/2} \left\{ {{a'}\over a} + {\dot a
\over a} {c^2 \over n^2} \dot{\cal R} \right\} \delta_i ^j \, ,\\
K^5 _5 &=& {1 \over c} \left( 1- {{c^2(t,{\cal R}(t))}\over
{n^2(t,{\cal R}(t))} } \right)^{-1/2} \left\{ {{b'}\over b} + {\dot b
\over b} {c^2 \over n^2} \dot{\cal R} \right\} \, ,\\
K_6 ^6 &=& -{c^2 \over n^2} \dot{\cal R}^2 K_0 ^0 \, ,
\end{eqnarray}
where all the coefficients take values on the brane.

\subsection{The Israel Conditions}

The energy-momentum tensor on the brane located at $z_0$ can be written as
\begin{equation}\label{em}
T^{(b)} _{MN} = {{\delta (z-z_0)} \over c} \left\{ (\rho +p) u_M u_N +
p \, \eta_{MN} \right\}\, .
\end{equation}
We also define a tensor $\hat T_{AB}$ as
\begin{equation}\label{tab}
\hat T_{AB} \equiv T_{AB} - {1 \over 4} T \eta_{AB} \, .
\end{equation}

The Israel junction conditions \cite{wisrael} are given by
\begin{equation}\label{israel}
[ K_{\mu\nu} ] = -\kappa_{(6)} ^2 \hat T_{\mu\nu} \, ,
\end{equation}
where the brackets stand for the jump across the brane and
$K_{\mu\nu}=e_\mu ^A e_\nu ^B K_{AB}$, where $e_\mu ^A$ form a basis
of the vector space tangent to the brane worldvolume.

The non-zero components of (\ref{tab}) are given by
\begin{eqnarray}\label{hatT}
\hat T_0^0 &=& -{{\gamma^2} \over c} \left\{ {{3\rho + 4 \, p} \over
    4} \right\} \, ,\\
\hat T_0 ^6 &=& \dot{\cal R} \hat T_0 ^0 \, , \quad \quad \quad \hat
    T_6 ^0 = - {c^2 \over n^2} \dot{\cal R} \hat T_0 ^0 \, , \\
\hat T_i ^j &=& {\rho \over {4c}} \delta_i ^j \, , \quad \quad \quad
    \hat T_5^5 = \hat T_i ^i \, , \\
\hat T_6 ^6 &=& -{c^2 \over n^2} \dot{\cal R}^2 \hat T_0 ^0 \, .
\end{eqnarray}

The left-hand side of (\ref{israel}) can be calculated taking into
account the mirror symmetry across the brane
\begin{equation}\label{mirror}
[K_{\mu\nu}] = K_{\mu\nu} (t,{\cal R}(t)^+) - K_{\mu\nu} (t,{\cal
  R}(t)^-) = -2K_{\mu\nu} (t,{\cal R}(t)) \, .
\end{equation}

At this point it is convenient to choose a specific bulk metric
of the form (\ref{metric}) satisfying six-dimensional Einstein
equations. This is given by
\begin{equation}\label{bhmetric}
ds^2 = - h(z) dt^2 + a^2(z) d\Sigma_k ^2 + h^{-1}
(z) dz^2 \, ,
\end{equation}
where
\begin{equation}\label{az}
a(z) = {z \over l} \, ,
\end{equation}
\begin{equation}\label{spacediff}
d\Sigma_k ^2 = {{dr^2} \over {1-kr^2}} + r^2 d\Omega_{(2)} ^2 +
(1-kr^2) dy^2 \, ,
\end{equation}
and
\begin{equation}\label{h} 
h(z) = k + {z^2 \over l^2} - {M \over z^3} + {Q^2 \over z^6} 
\end{equation}
with $l^{-2} \propto -\Lambda$ ($\Lambda$ being the cosmological
constant, which can be positive or negative) and $M$ and $Q^2$ are
constants. 

We should stress that $dy$ can be written as $dy=R_c \, d\varphi$, with
$\varphi$ an angle with the usual periodicity $2\pi$. Thus, the
coordinate $y$ is compactified under some mechanism such that its
radius of compactification $R_c$ is small enough to evade experimental
detection \cite{KK,arkani}. This makes the local observers have the picture
of living on an effective three-dimensional brane.

Metric (\ref{bhmetric}) contains a singularity located at
$z=0$. It is valid on the $z <{\cal R}(t)$ parts of surfaces of constant
$t$ and its reflection, by the $Z_2$ orbifold symmetry, is valid on the
$z >{\cal R}(t)$ parts. If $M=0$ and 
$Q^2=0$, then (\ref{bhmetric}) is simply the metric of de Sitter or anti
de Sitter spacetime according to the sign of $l^2$. 

With this Ansatz the Israel conditions (\ref{israel}) reduce to only
two equations, which read as
\begin{eqnarray}\label{brem}
\ddot{\cal R} + {1\over 2} {{h'}\over h^3} \dot{\cal R}^4 - 3
{{h'}\over h} \dot{\cal R}^2 + {1 \over 2} h \, h' &=& -\kappa_{(6)} ^2
\left( {{3\rho + 4 \, p}\over 8} \right) h^2 \left( 1- {{\dot{\cal
        R}^2} \over h^2} \right)^{3/2} \nonumber \\ \\
{{a'} \over a} + {{\dot{\cal R}}\over h^2} {{\dot a}\over a} &=&
\kappa_{(6)} ^2 {\rho\over 8} \left( 1- {{\dot{\cal
        R}^2} \over h^2} \right)^{1/2} \, , \nonumber
\end{eqnarray}
where again all the metric coefficients must be evaluated on the
brane. System (\ref{brem}) describes the full nonlinear dynamics
of the brane embedded in the static bulk (\ref{bhmetric}).

\subsection{The Geodesic Equation and the Time Delay}

We consider two points on the brane $r_A$ and $r_B$. In general there
are more than one null geodesic connecting these points in the $1+5$
spacetime. The trajectories of photons must be on the brane and those
of gravitons may be outside. The graviton path is defined equating
(\ref{bhmetric}) to zero.
Since we are looking for a path that minimizes $t$ when the final
point $r_B$ is on the brane, the problem reduces to an
Euler-Lagrange problem \cite{acmfw}.
Then as in \cite{acm} the shortest graviton path is given by
\begin{equation}\label{geo}
\ddot {\cal R}_g + \left( {1 \over {\cal R}_g} - {3 \over 2}
  {{h'}\over h}\right) \dot {\cal R}_g^2 + {1\over 2} h \, h' - {h^2
  \over {\cal R}_g} = 0 \; . 
\end{equation}

We can also calculate the time delay between the photon travelling on
the brane and the gravitons travelling in the bulk. Since both signals
cover the same distance \cite{ac},  
\begin{equation}\label{timedelay}
\int_0 ^{\tau_f+\Delta\tau} {{d\tau_\gamma}\over {{\cal R}(\tau_\gamma)}} =
\int_0 ^{t_f} {{dt_g}\over
{{\cal R}_g(t_g)}} \sqrt{h({\cal R}_g) - {{\dot {\cal R}_g(t)^2}\over {h({\cal R}_g)}}} \, ,
\end{equation}
the difference between photon and graviton flight times measured by an
observer on the brane can
approximately be written in terms of the bulk time $t$ as \cite{acm}
\begin{equation}\label{delay} 
\Delta\tau \simeq {\cal R}(t_f) \int_0 ^{t_f} dt \left({{1}\over
{{\cal R}_g(t)}} 
\sqrt{h({\cal R}_g)- {{\dot {\cal R}_g(t)^2}\over {h({\cal R}_g)}}} -
{1 \over {{\cal R}(t)}} 
{{d\tau}\over{dt}} \right) \, .
\end{equation}

It is also interesting to look at the ratio between the horizons
subtended by the photons travelling on the brane and the gravitons
travelling in the bulk. This ratio uses the same quantities previously
quoted for the time delay,
\begin{equation}\label{ratio}
{g \over \gamma} = { {\int_0 ^{t_f} {{dt}\over {{\cal R}_g(t)}}
\sqrt{h({\cal R}_g)- {{\dot {\cal R}_g(t)^2}\over {h({\cal R}_g)}}}}
\over {\int_0 ^{t_f} {{dt}\over {{\cal R}(t)}} {{d\tau}\over{dt}}} }
\, . 
\end{equation}

\section{Brane Fluctuations}

In this section we study perturbatively the brane behaviour at a fixed
point. We define the background solution as the case where the brane
position is frozen, {\it i.e.}
\begin{equation}\label{frozen}
z = {\cal R} (t) = \bar{\cal R} = const.
\end{equation}
Then the system (\ref{brem}) reduces to
\begin{eqnarray}
{1 \over 2} {{h'}\over h} &=& - {{\kappa_{(6)} ^2} \over 8} (3 \bar\rho
+4 \bar p) \label{linearh} \\ 
{{a'} \over a} &=& {{\kappa_{(6)} ^2} \over 8} \bar\rho \, . \label{lineara}
\end{eqnarray}

The equations of motion for the brane fluctuations can be obtained
linearizing the exact equations of motion (\ref{brem}) about the
``equilibrium'' configuration. They read as
\begin{eqnarray}\label{system}
{{\ddot{\delta{\cal R}}} \over h^2} + {1 \over 2} \delta \left(
  {{h'}\over h} \right) &=& - {{\kappa_{(6)} ^2} \over 8} (3 \,
  \delta\rho + 4 \, \delta p) \nonumber \\ \\
\delta \left( {{a'}\over a} \right) &=& {{\kappa_{(6)} ^2} \over 8}
  \delta\rho \, . \nonumber
\end{eqnarray}
From the explicit expressions of $h$ (with $Q=0$) and $a$ we can find
\begin{eqnarray}
\delta \left( {{a'}\over a} \right) &=& - {{\kappa_{(6)} ^4} \over
  {64}} \bar\rho ^2 \delta{\cal R} \equiv m_a ^2 \delta{\cal R}
  \label{ma} \\
{1 \over 2} \delta \left( {{h'}\over h} \right) &=& \left\{ -
  {{\kappa_{(6)} ^4} \over {32}} (3\, \bar\rho + 4\, \bar p)^2 + {{{1
  \over l^2} -6 {M \over {\bar{\cal R}^5}}} \over {k + {{\bar{\cal
  R}^2}\over l^2} - {M \over {\bar{\cal R}^3}}}} \right\} \delta{\cal
  R} \equiv m_h ^2 \delta{\cal R} \, . \label{mh}
\end{eqnarray}
As we can see from (\ref{system}), the brane fluctuations are bound to
the matter fluctuations. However, when we have the case of adiabatic
matter fluctuations
\begin{equation}\label{adiabatic}
\delta p = v_s ^2 \delta\rho \, ,
\end{equation}
we can derive an equation for the brane fluctuations with an
appropriate linear combination of (\ref{system})
\begin{equation}\label{fluctuation}
{{\ddot{\delta{\cal R}}}\over h^2} + \left( m_h ^2 + (3+ 4 \, v_s ^2)
  m_a ^2 \right) \delta{\cal R} = 0 \, ,
\end{equation}
while the matter fluctuations will be related to the brane ones by
\begin{equation}\label{rhoflu}
{{\kappa_{(6)} ^2} \over 8} \delta\rho = m_a ^2 \delta{\cal R} \, .
\end{equation}
Assuming an equation of state $p=\omega\rho$, (\ref{fluctuation})
finally reads
\begin{equation}\label{fluc}
{{\ddot{\delta{\cal R}}}\over h^2} + \left\{ {{{1
  \over l^2} -6 {M \over {\bar{\cal R}^5}}} \over {k + {{\bar{\cal
  R}^2}\over l^2} - {M \over {\bar{\cal R}^3}}}} - (7+8\omega)
  (3+4\omega) {{\kappa_{(6)} ^4} \over {64}} \bar\rho ^2 \right\}
  \delta{\cal R} =0 \, .
\end{equation}
The constant $M$ can be expressed in terms of $\bar{\cal R}$ and
$\omega$ from the Israel conditions for a static brane \cite{accm} as
\begin{equation}\label{Mstatic}
M = {{2 \bar{\cal R}^3 } \over {8\omega + 3}} \left\{ (4\omega +3) k +
  4 (\omega+1) {{\bar{\cal R}^2}\over l^2} \right\} \, .
\end{equation}

In the case of a flat domain wall ($k=0$, $\omega =-1$), the term inside braces
in (\ref{fluc}) can be interpreted as an effective cosmological
constant on the brane, 
\begin{equation}\label{eff}
{1\over {\bar{\cal R}^2}} - {{\kappa_{(6)} ^4} \over {64}} \bar\rho ^2
\equiv -\Lambda_{eff} \, ,
\end{equation}
which vanishes when we take into account (\ref{lineara}). This
corresponds to the Randall-Sundrum condition \cite{rs1,rs2} in six
dimensions. Then, the effective cosmological constant on the brane is
zero.

Therefore, we are left with the equation of motion for a free scalar
field as in five dimensions \cite{radion,cgr}. In this case the brane
is at rest and the equilibrium position $\bar{\cal R}$ can be chosen
arbitrarily. However, a small departure from this position results in
a runaway behaviour.

This result is compatible with the numerical solution of the full
nonlinear system (\ref{brem}) as we will see in the next section.

\begin{figure*}[h!]
\begin{center}
\leavevmode
\begin{eqnarray}
\epsfxsize= 3.7truecm\rotatebox{-90}
{\epsfbox{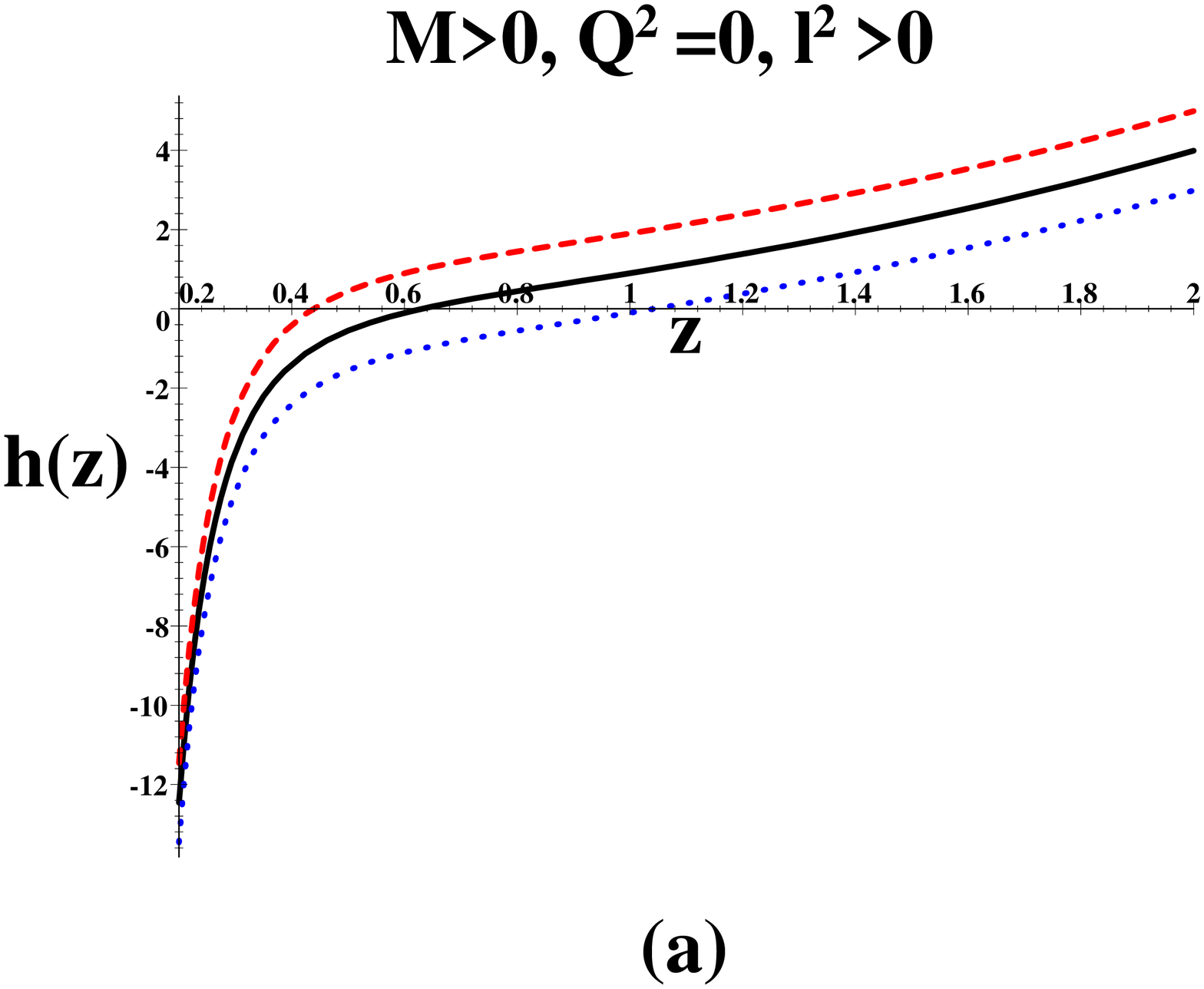}}\nonumber
\epsfxsize= 3.7truecm\rotatebox{-90}
{\epsfbox{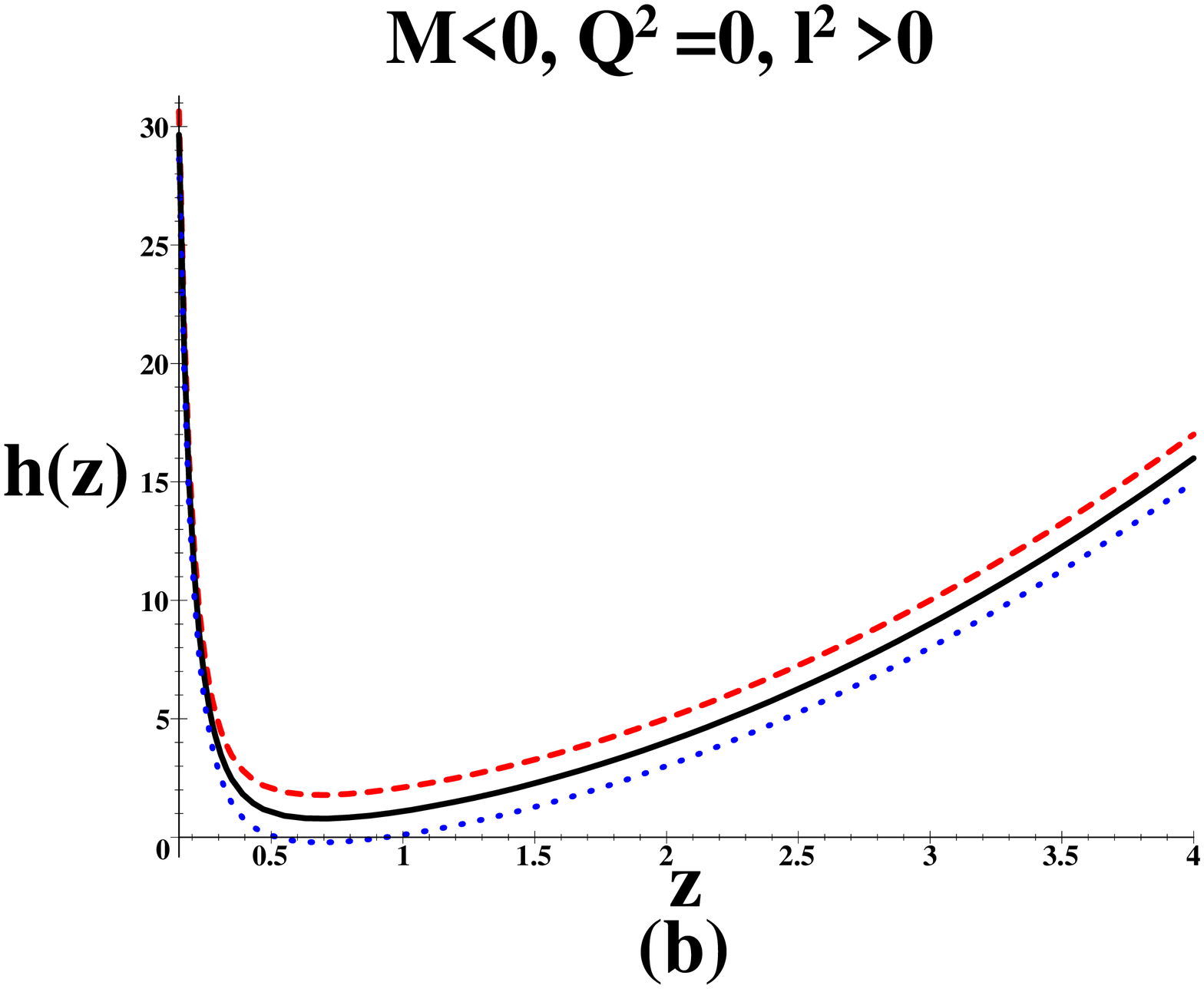}}\nonumber
\epsfxsize= 3.7truecm\rotatebox{-90}
{\epsfbox{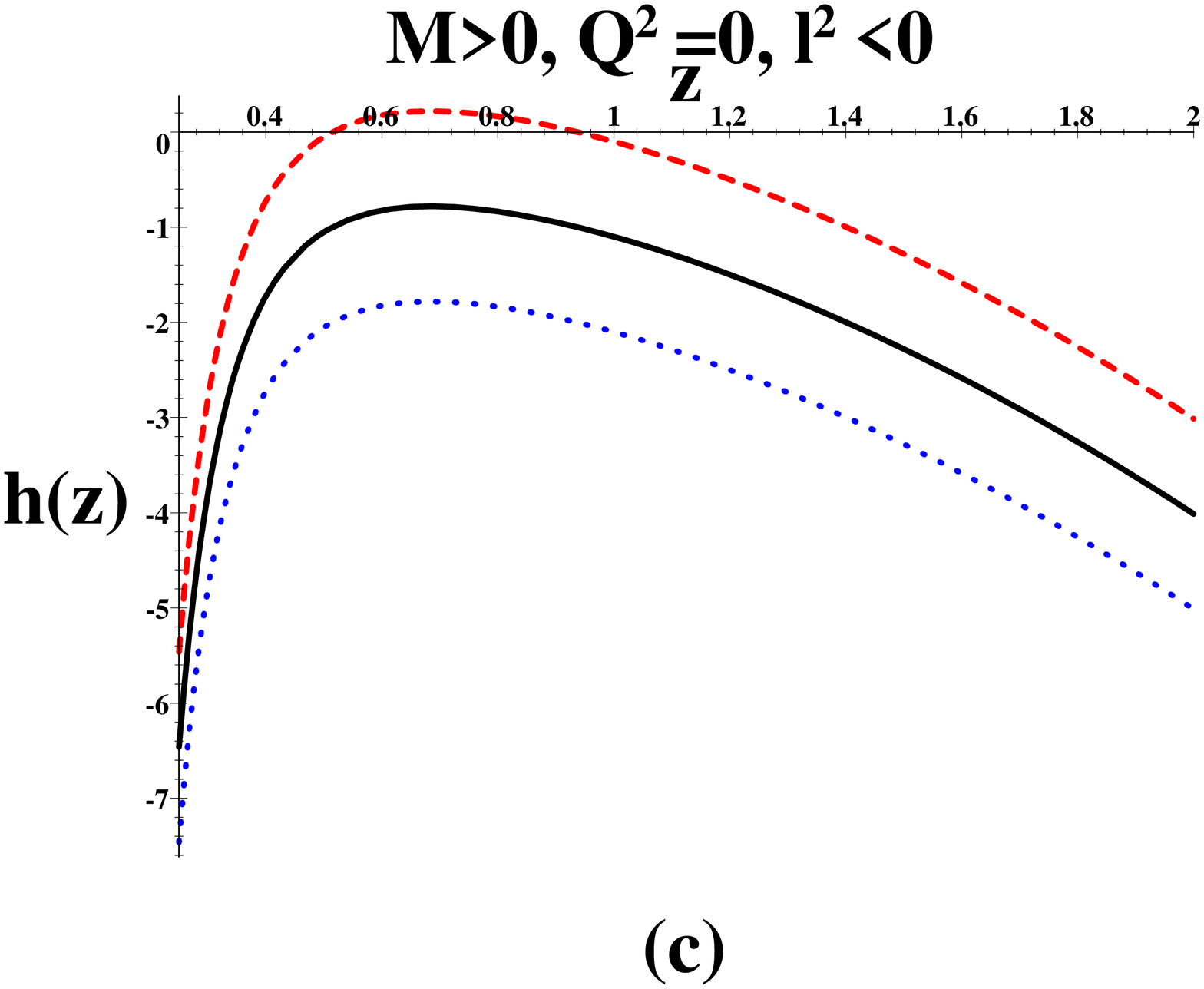}}\nonumber\\
\epsfxsize= 3.7truecm\rotatebox{-90}
{\epsfbox{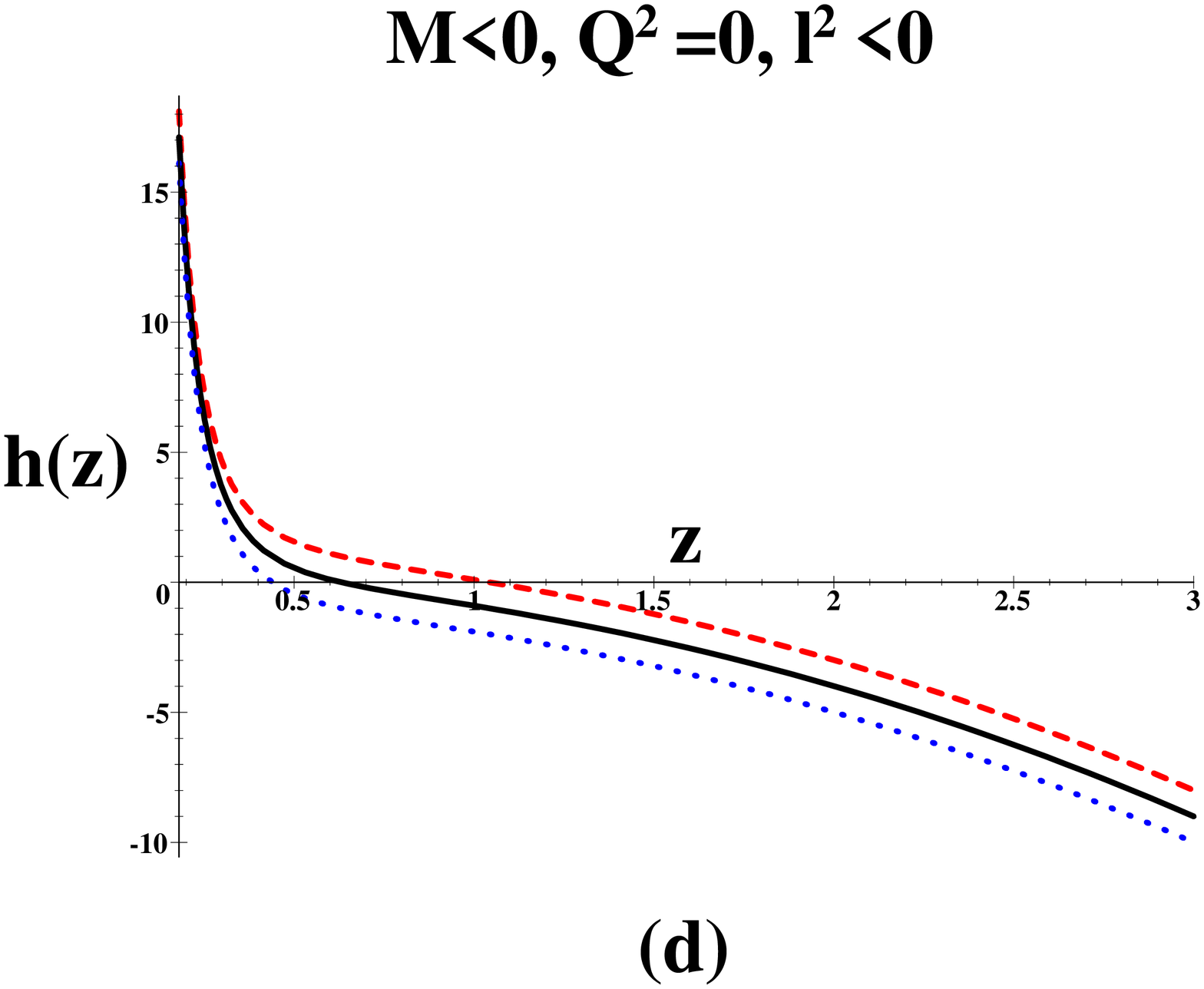}}\nonumber
\epsfxsize= 3.7truecm\rotatebox{-90}
{\epsfbox{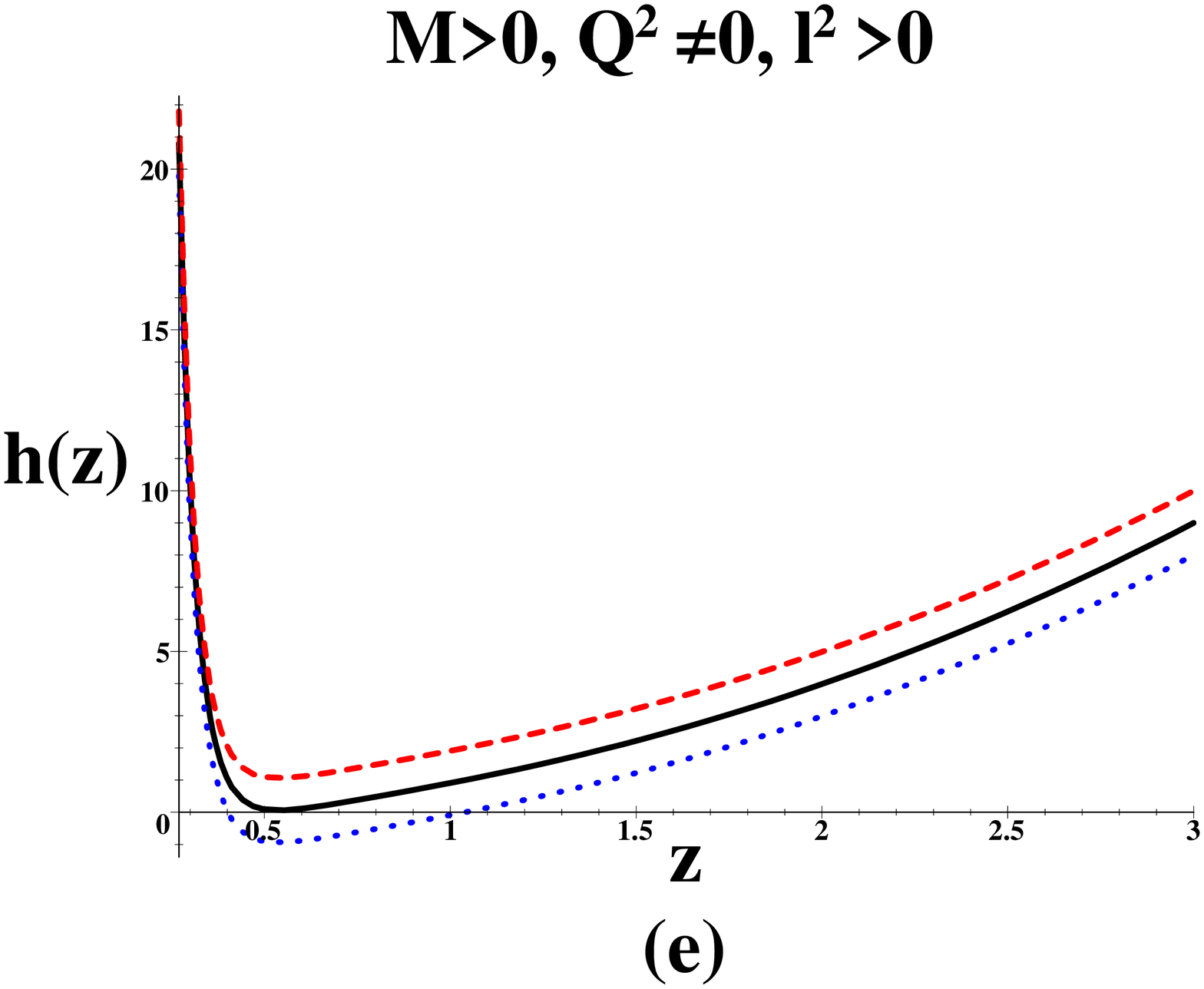}}\nonumber
\epsfxsize= 3.7truecm\rotatebox{-90}
{\epsfbox{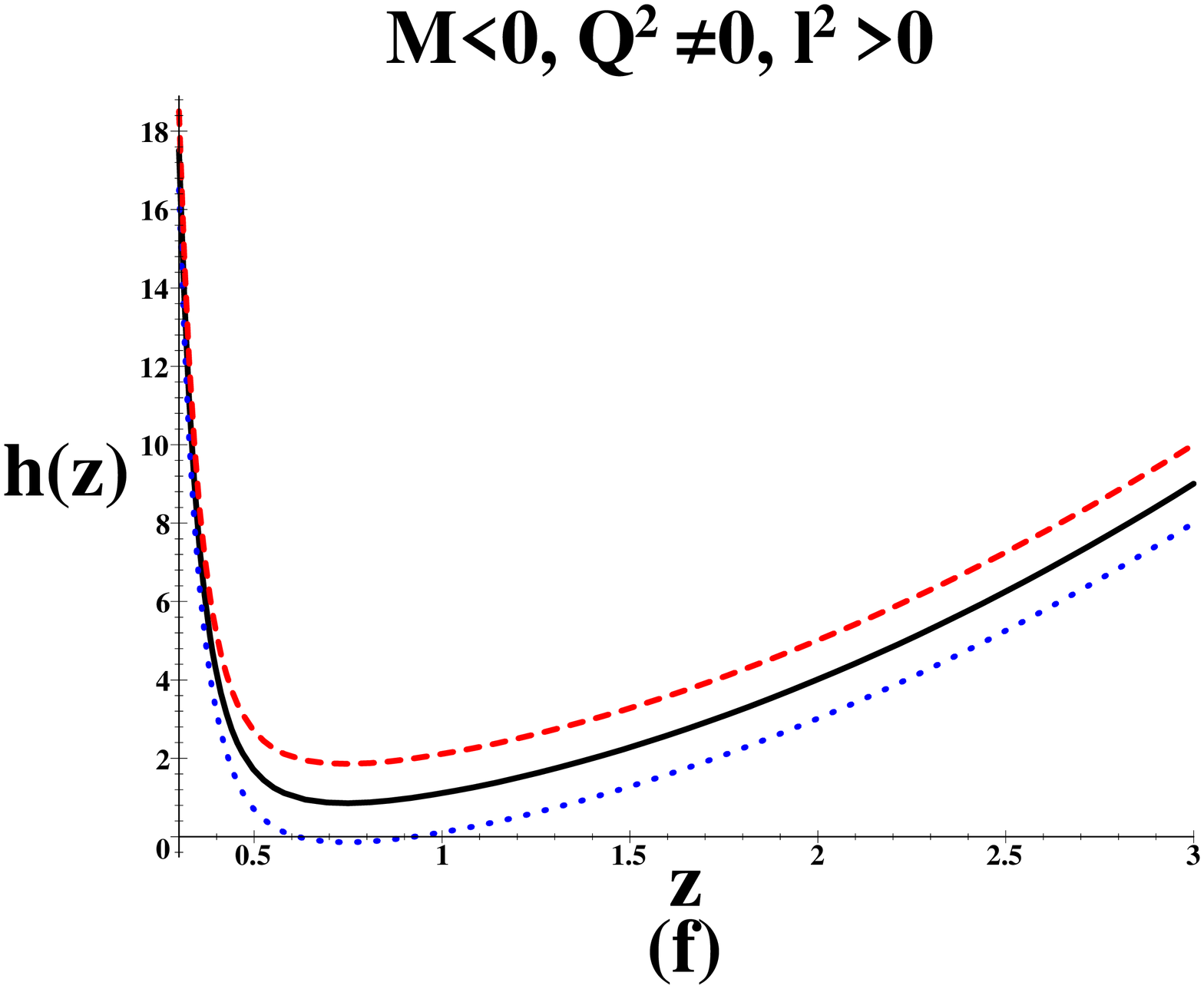}}\nonumber\\
\epsfxsize= 3.7truecm\rotatebox{-90}
{\epsfbox{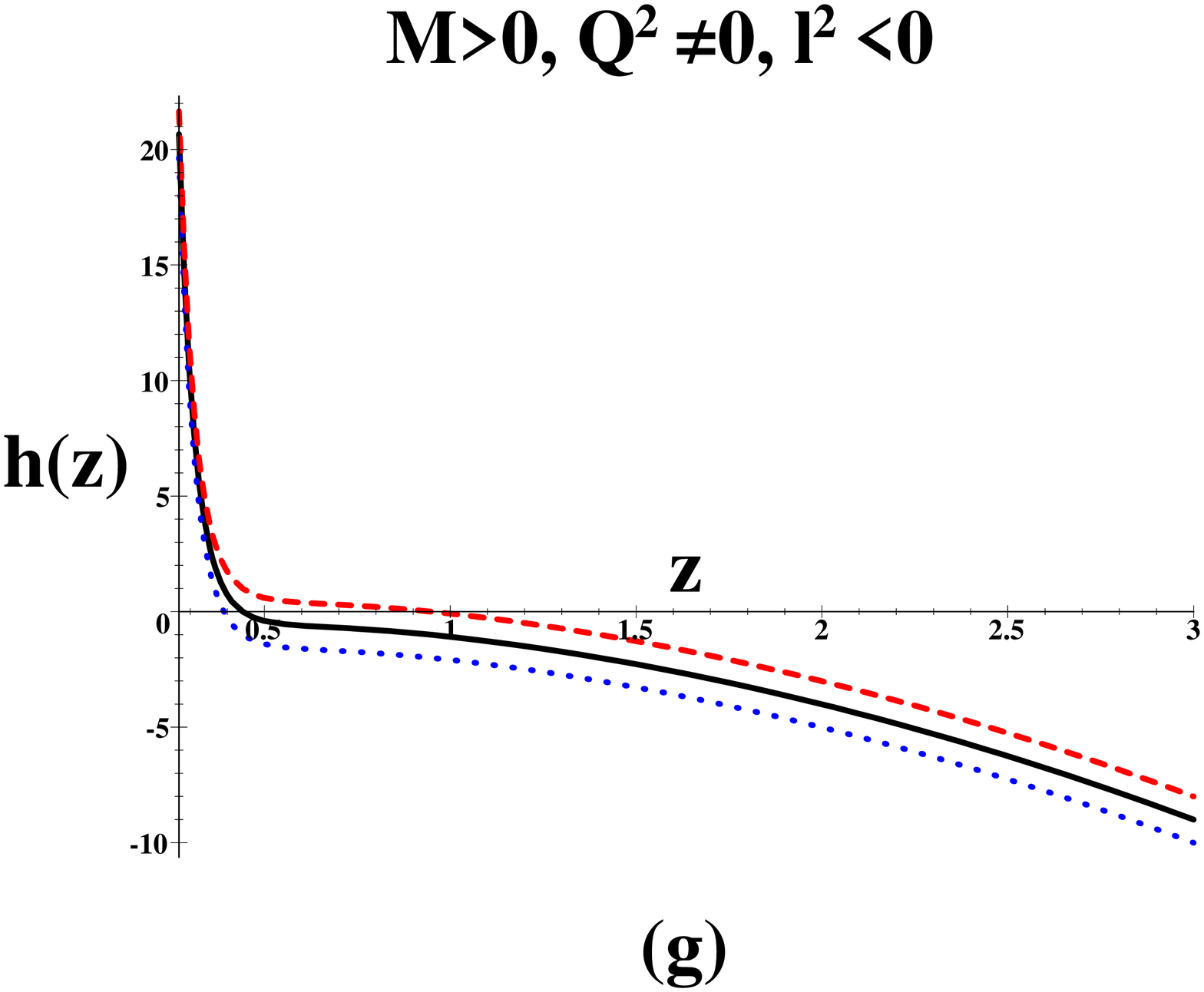}}\nonumber
\epsfxsize= 3.7truecm\rotatebox{-90}
{\epsfbox{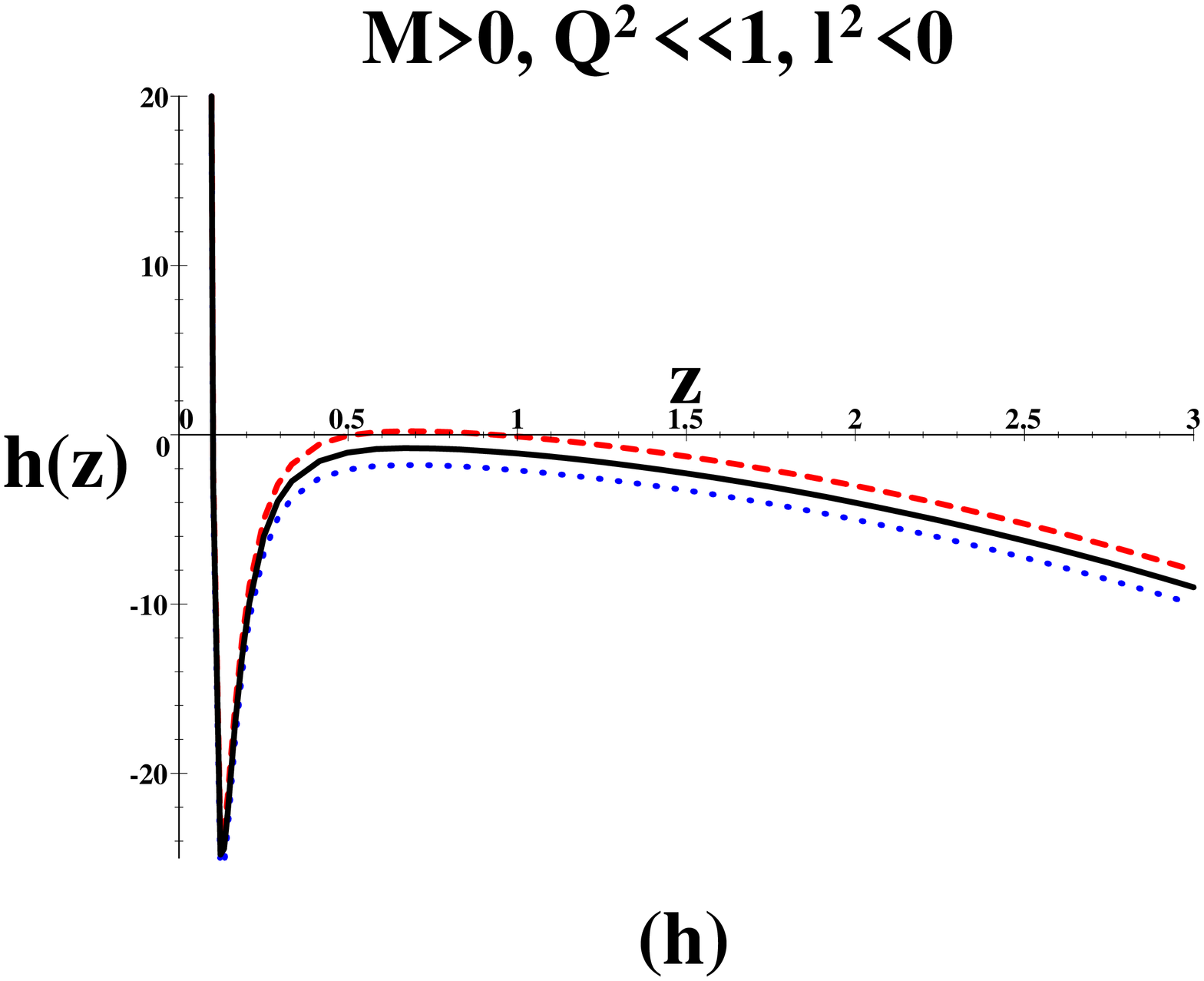}}\nonumber
\epsfxsize= 3.7truecm\rotatebox{-90}
{\epsfbox{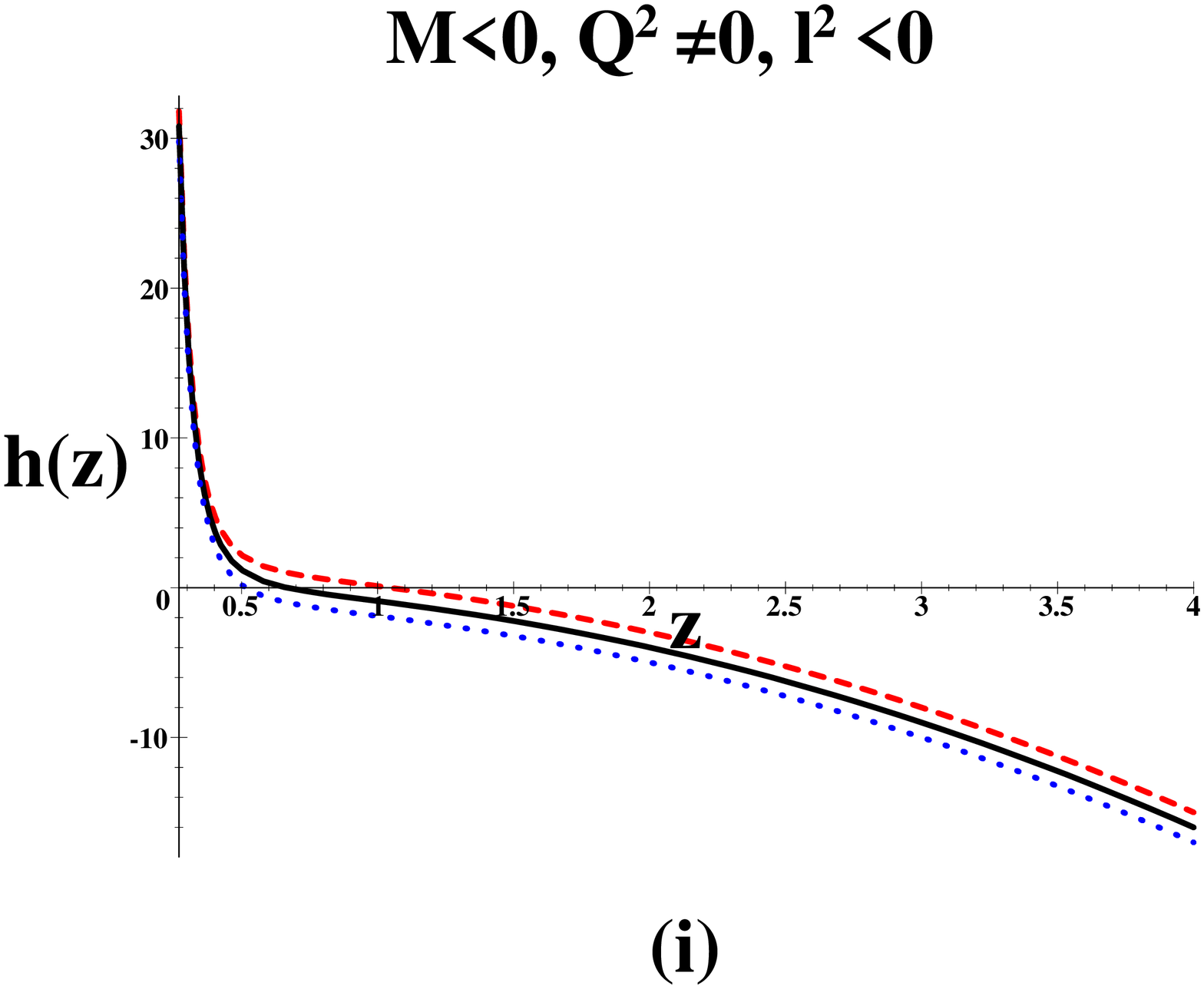}}\nonumber
\end{eqnarray}
\caption{$h(z)$ for all the possible combinations of $M$, $Q^2$ and
  $l^2$, including $k=1$ (dashed lines), $k=0$ (solid lines) and $k=-1$
  (dotted lines) cases.} 
\label{hh}
\end{center}
\end{figure*}

\section{Nonlinear Brane Dynamics}

The system (\ref{brem}) describing the brane dynamics can be
numerically solved for several combinations of $M$, $Q^2$, $k$ and
$l^2$. When $M$, $Q^2$ and $k$ vanish, the solution for ${\cal R}(t)$
is a constant or a linear function in $t$ depending on the given initial
condition for $\dot{\cal R}$, and we verify the behaviour found through
brane fluctuations.

For $M$ and $Q^2$ non-zero we have solved (\ref{brem}) in the typical
cases of a domain wall ($\omega=-1$), matter ($\omega=0$) and
radiation ($\omega=1/3$) dominated branes. In figure \ref{hh} we show
all the possibilities for the bulk metric according to the form that
$h(z)$ takes due to $M$, $Q^2$, $k$ and $l^2$  
combinations, which we study in this section. Some of our results are
also illustrated in what follows together with the solution of the geodesic
equation (\ref{geo}) in order to verify the possibility of having
shortcuts. We have also calculated the time delays and the ratio
between graviton and photon horizons for the examples of
shortcuts appearing in this paper, these are shown in Table \ref{td}
together with the graviton bulk flight time and its corresponding
brane time according to the equation (\ref{dtdtau}).

We have classified all cases according to the sign of the $M$ 
parameter and to whether we are in dS or AdS bulks. Moreover, we
studied the zero charge black hole as well as the
Reissner-Nordstr\"om-type solutions, namely eight cases.

\begin{table}[htb!]
\begin{tabular}{|c|c|c|c|c|c|c|c|}\hline\hline
$M$ &$k$ &$l^2$ &Bulk   &DW&MDB&RDB&Geodesic\\ \hline \hline  
 +  & 1   & +   &{\footnotesize AdS-Schwarzschild} & {\footnotesize
 $\rightarrow r_H$/{\it G}}&{\footnotesize $\rightarrow 
 r_H$}&{\footnotesize $\rightarrow r_H$}&{\footnotesize $\rightarrow
 {r_H} ^*$/{\it G}}\\  
 +  &0,-1 &+    &{\footnotesize AdS-top. black hole}& {\footnotesize
 $\rightarrow r_H$}&{\footnotesize $\rightarrow 
 r_H$}&{\footnotesize $\rightarrow r_H$}&{\footnotesize $\rightarrow
 {r_H} ^*$} \\ 
 -  & 0,1 & +   &{\footnotesize AdS-naked singularity}&
 {\footnotesize {\it G}}&{\footnotesize $\rightarrow 0$}&{\footnotesize
 $\rightarrow  0$} &{\footnotesize {\it G}$^*$/$\rightarrow 0$}\\  
 -  & -1  & +   &{\footnotesize AdS-top. black hole}& {\footnotesize
 $\rightarrow r_H$}&{\footnotesize $\rightarrow 
 r_H$}&{\footnotesize $\rightarrow r_H$}&{\footnotesize $\rightarrow
 r_H$} \\  
 +  & 1   & -   &{\footnotesize dS-Schwarzschild}  & {\footnotesize
 $\rightarrow r_H$/$r_c$}&{\footnotesize $\rightarrow
 r_H$/$r_c$}&{\footnotesize $\rightarrow r_H$/$r_c$} 
 &{\footnotesize $\rightarrow r_H$/${r_c}^{*\dagger}$}\\   
 +  & 0,-1& -   &{\footnotesize dS-cosm. singularity}&
 {\footnotesize no solution} 
 &{\footnotesize no solution}&{\footnotesize no
 solution}&{\footnotesize no solution}\\   
 -  &0,$\pm$1& -&{\footnotesize dS-naked singularity}&
 {\footnotesize $\rightarrow r_c$}&{\footnotesize $\rightarrow
 0$/$r_c$}&{\footnotesize $\rightarrow 
 0$/$r_c$}&{\footnotesize $\rightarrow 0$/${r_c}^{*\dagger}$}\\ 
\hline \hline 
\end{tabular}
\caption{Scale factor and geodesics evolution (uncharged case). The
arrow indicates the behaviour tendency, which depends on the initial
conditions. {\it G} means growing behaviour. The $*$ or the $\dagger$ 
in the last column indicates the possibility of shortcuts for the
matter and radiation-dominated branes or the domain wall, respectively.}
\label{sum1}
\end{table}

\begin{figure*}[htb!]
\begin{center}
\leavevmode
\begin{eqnarray}
\epsfxsize= 6.0truecm\rotatebox{-90}
{\epsfbox{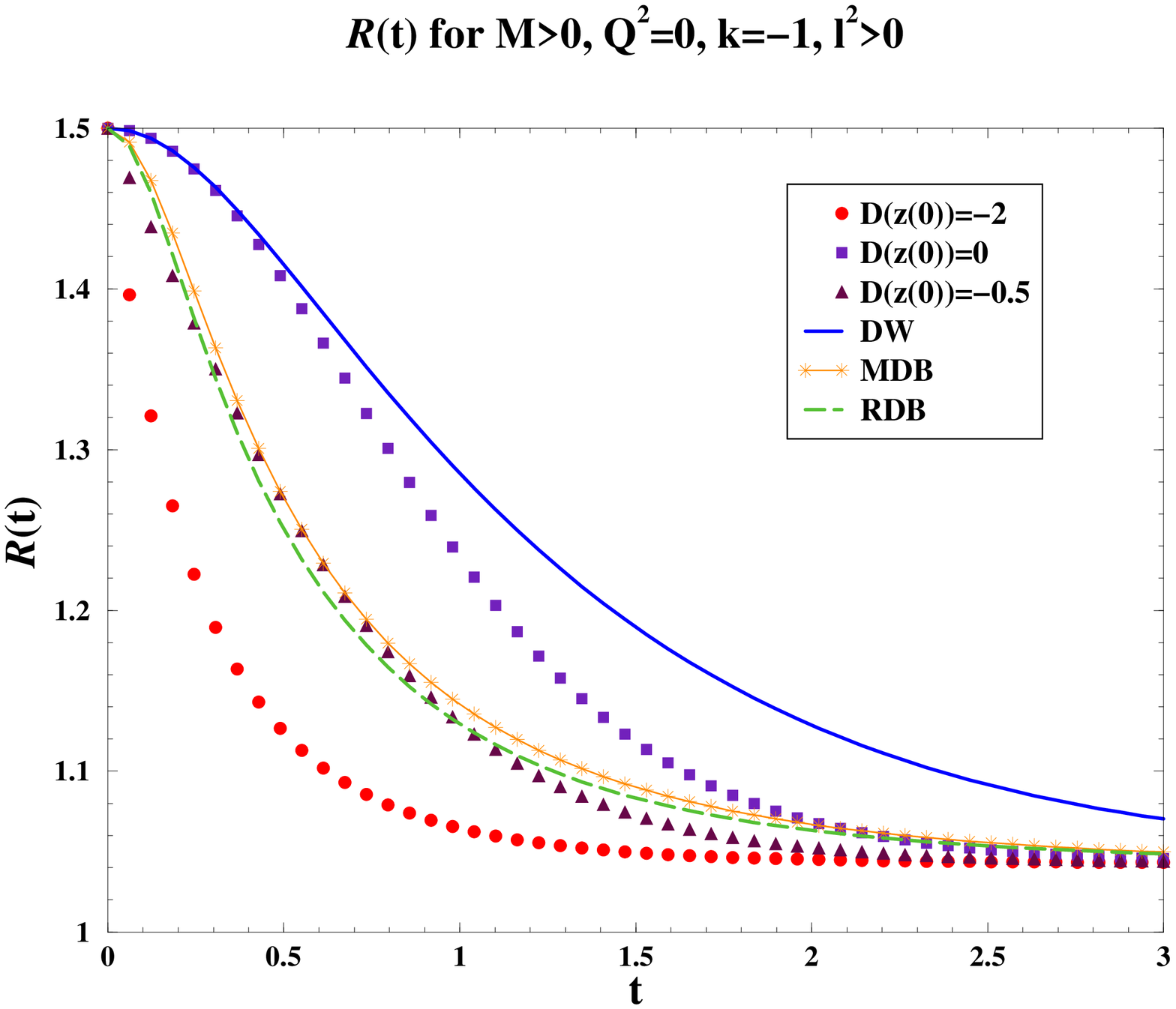}}\nonumber
\epsfxsize= 6.0truecm\rotatebox{-90}
{\epsfbox{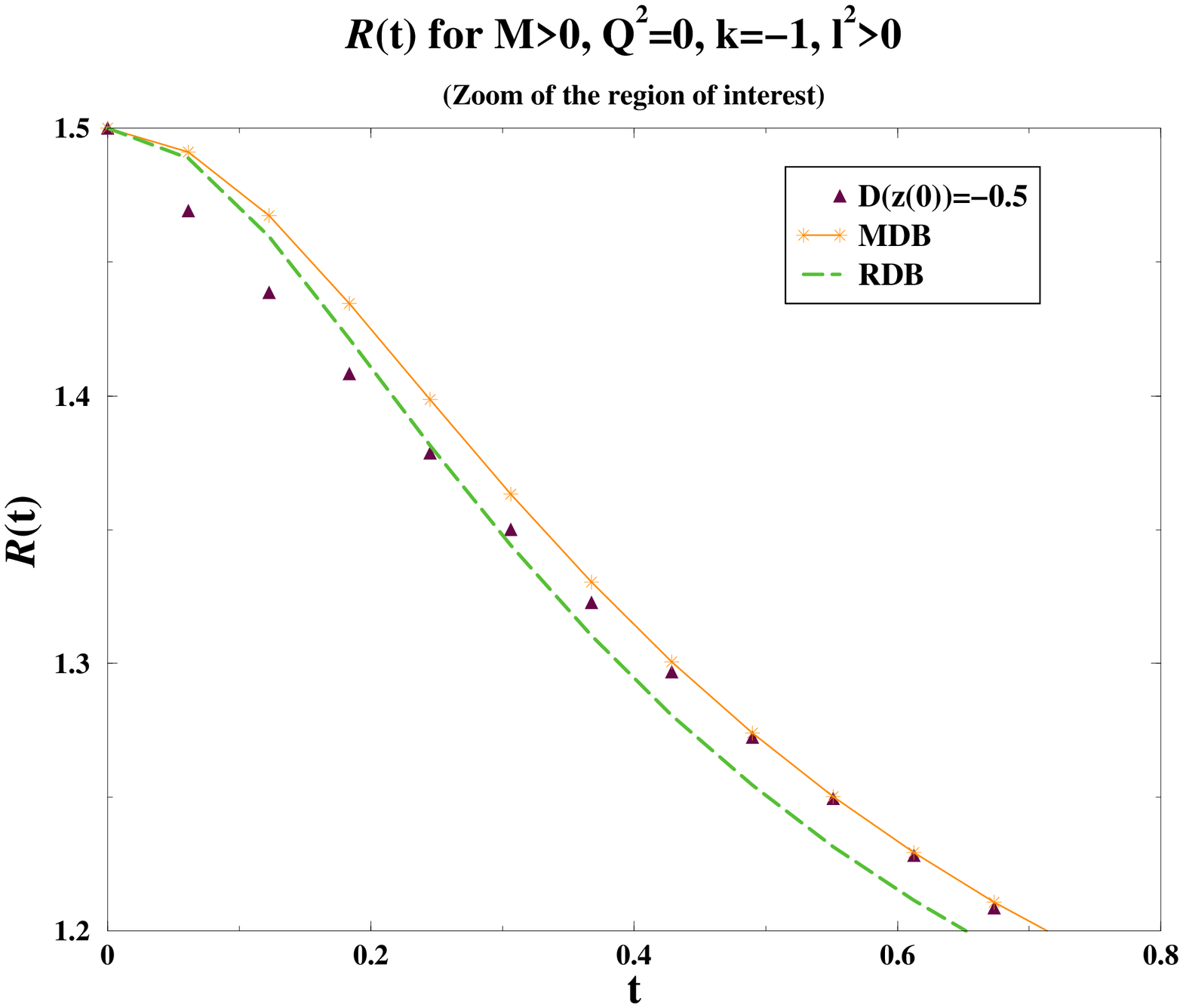}}\nonumber
\end{eqnarray}
\caption{Scale factor evolution for domain wall, matter and
radiation-dominated branes and geodesics when $M>0$, $Q^2=0$ and
$l^2>0$.}  
\label{sf1}
\end{center}
\end{figure*}
\begin{figure*}[htb!]
\begin{center}
\leavevmode
\begin{eqnarray}
\epsfxsize= 6.0truecm\rotatebox{-90}
{\epsfbox{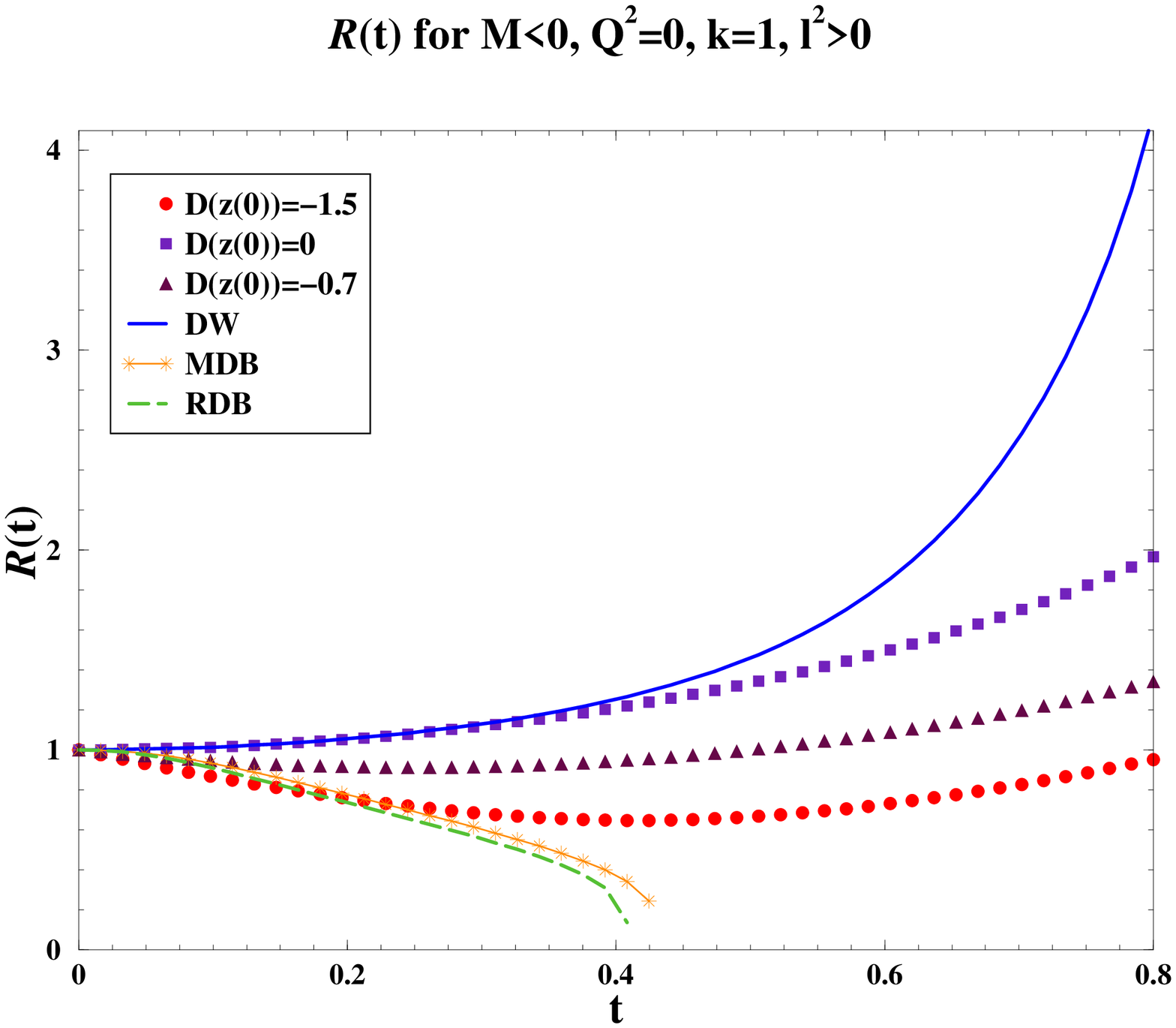}}\nonumber
\epsfxsize= 6.0truecm\rotatebox{-90}
{\epsfbox{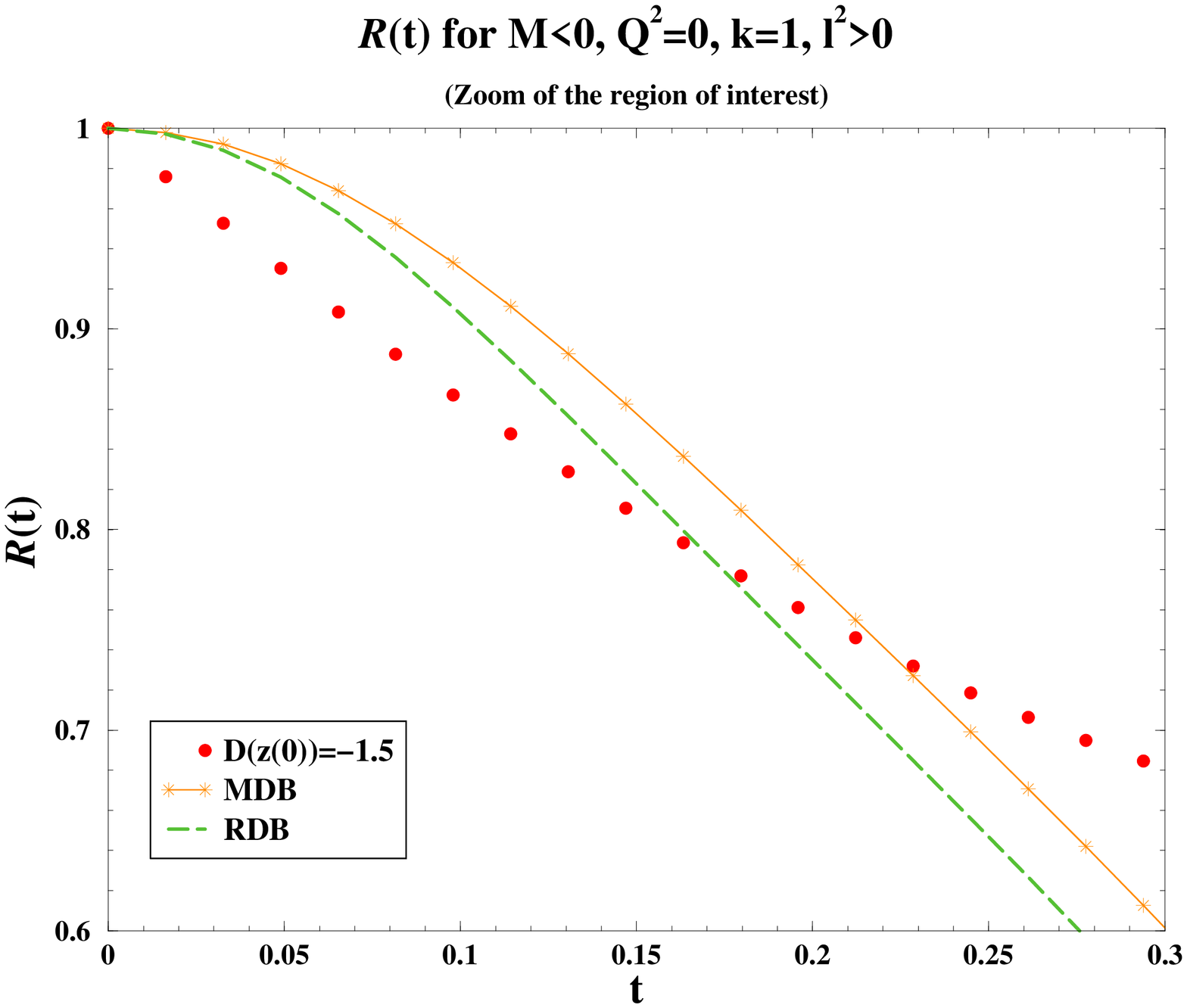}}\nonumber
\end{eqnarray}
\caption{Scale factor evolution for domain wall, matter and
radiation-dominated branes and geodesics when $M<0$, $Q^2=0$ and
$l^2>0$.}  
\label{sf2}
\end{center}
\end{figure*}
\begin{figure*}[htb!]
\begin{center}
\leavevmode
\begin{eqnarray}
\epsfxsize= 6.0truecm\rotatebox{-90}
{\epsfbox{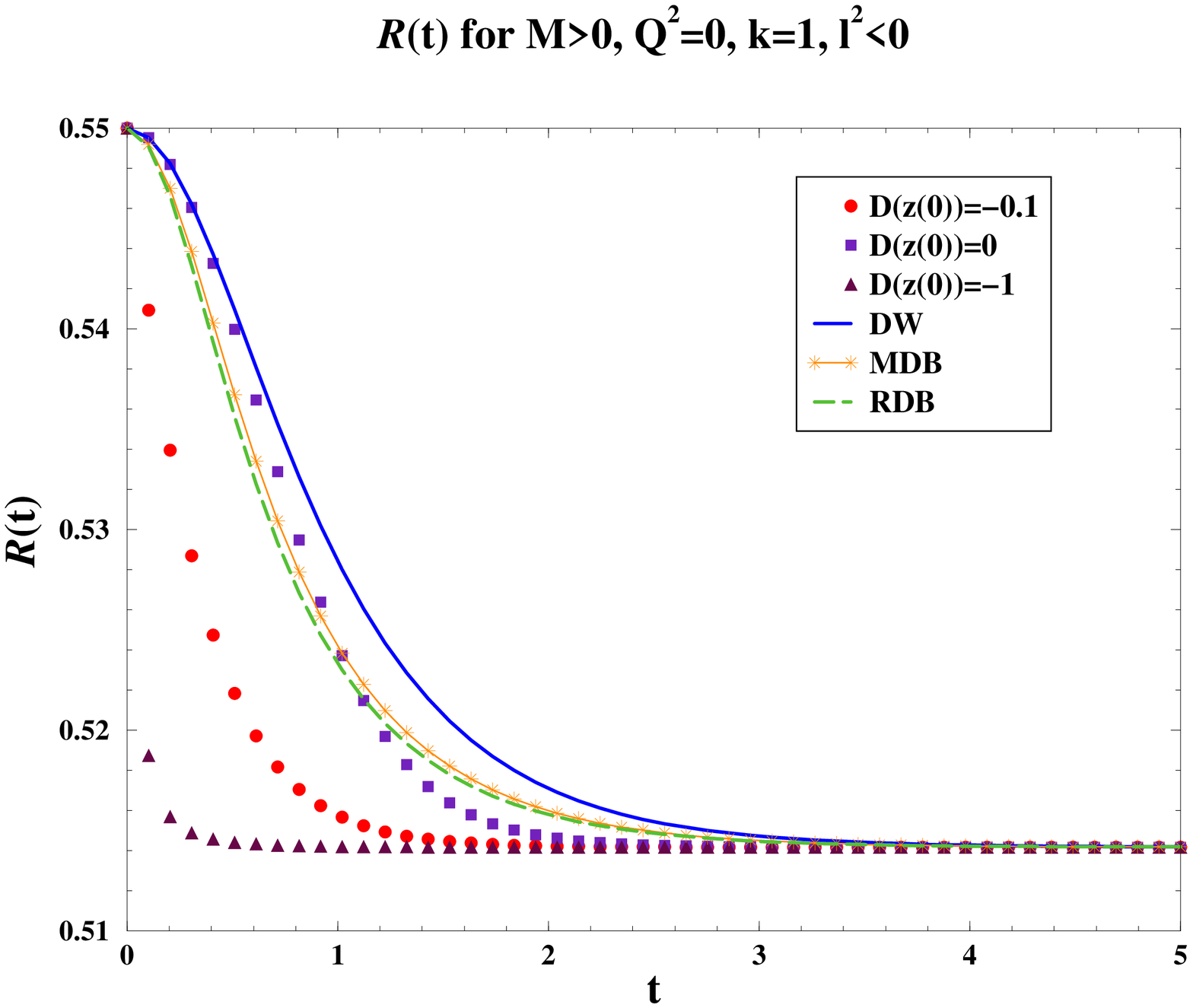}}\nonumber
\epsfxsize= 6.0truecm\rotatebox{-90}
{\epsfbox{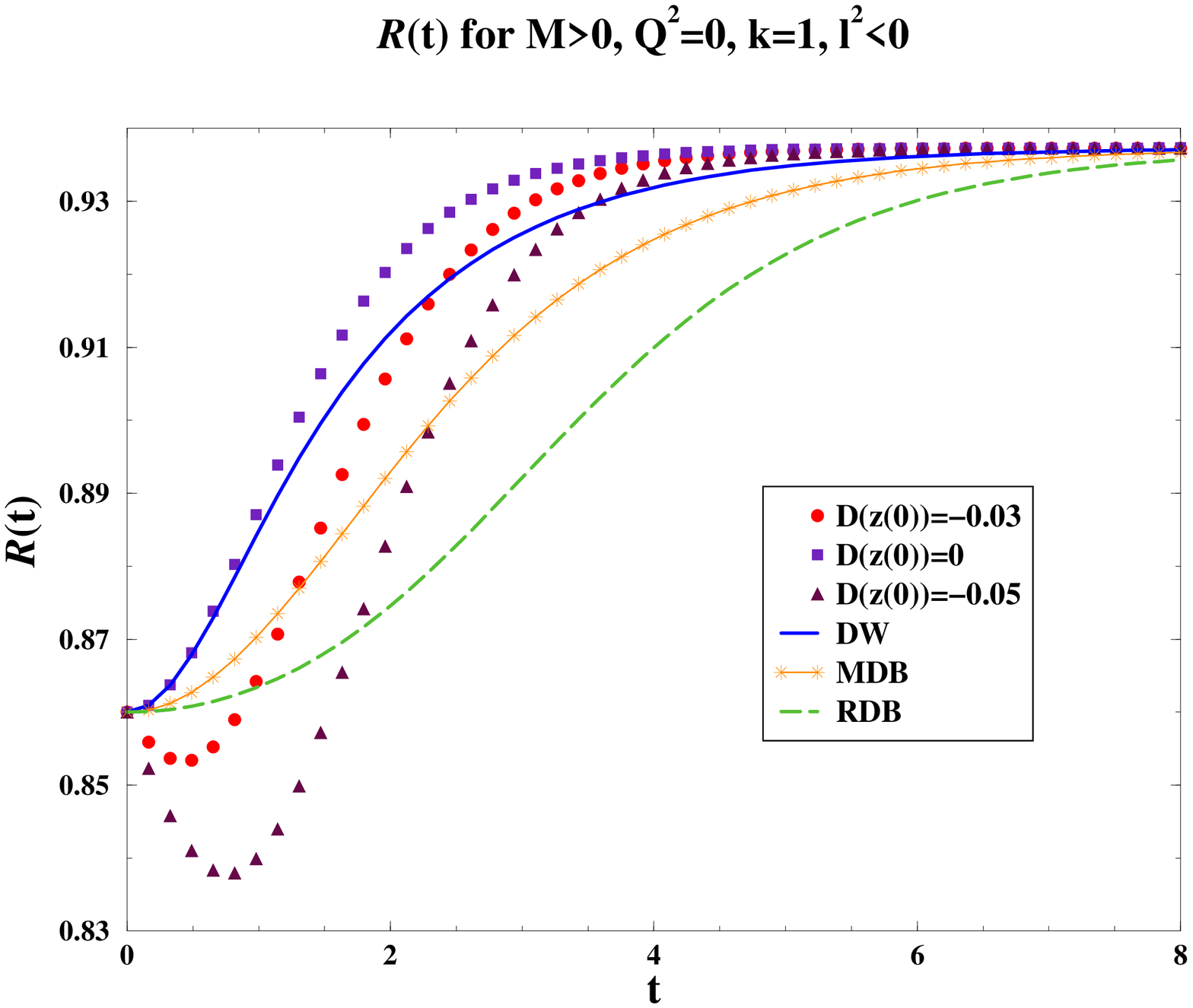}}\nonumber
\end{eqnarray}
\caption{Scale factor evolution for domain wall, matter and
radiation-dominated branes and geodesics when $M>0$, $Q^2=0$ and
$l^2<0$.}  
\label{sf3}
\end{center}
\end{figure*}
\begin{figure*}[htb!]
\begin{center}
\leavevmode
\begin{eqnarray}
\epsfxsize= 6.0truecm\rotatebox{-90}
{\epsfbox{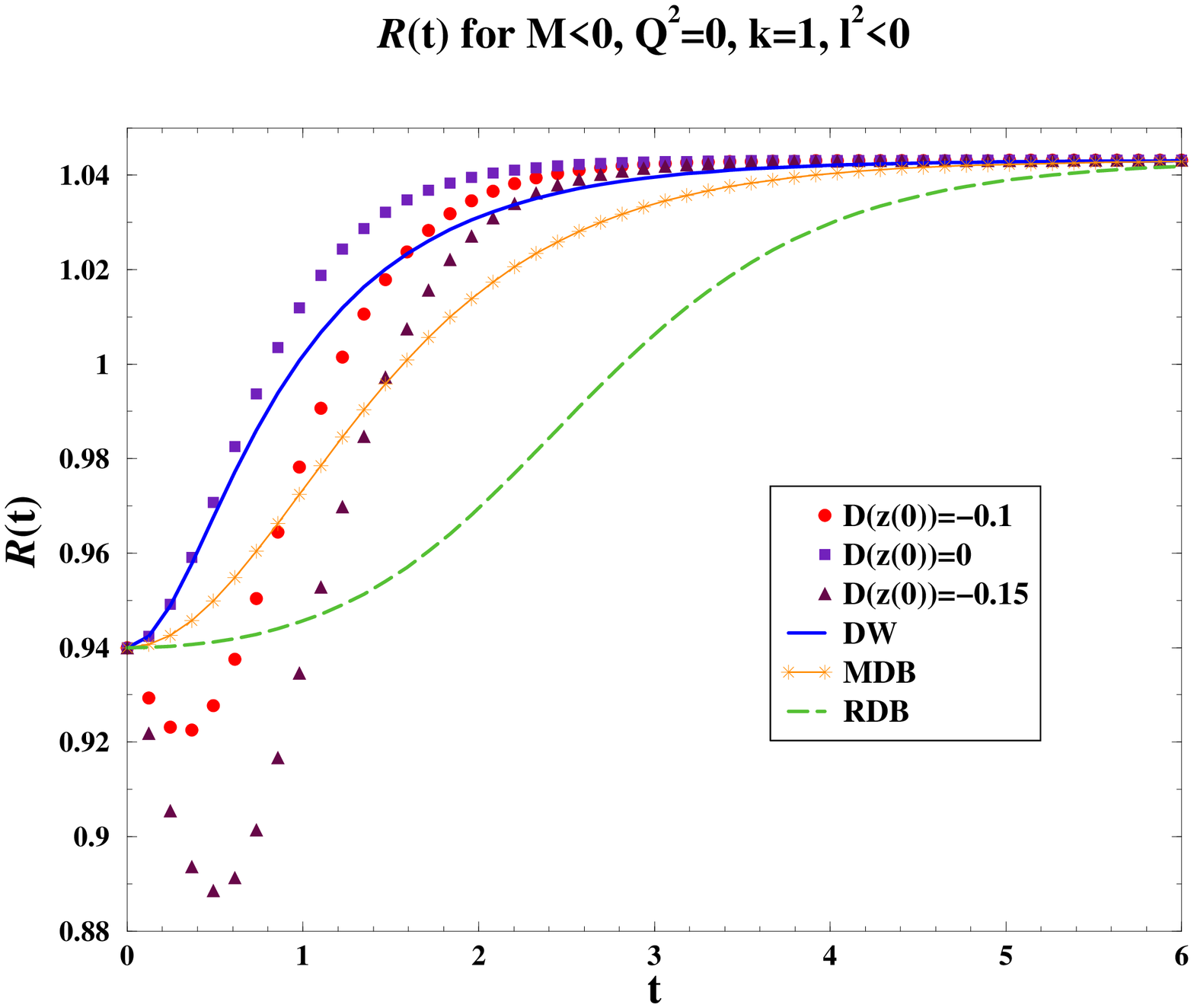}}\nonumber
\epsfxsize= 6.0truecm\rotatebox{-90}
{\epsfbox{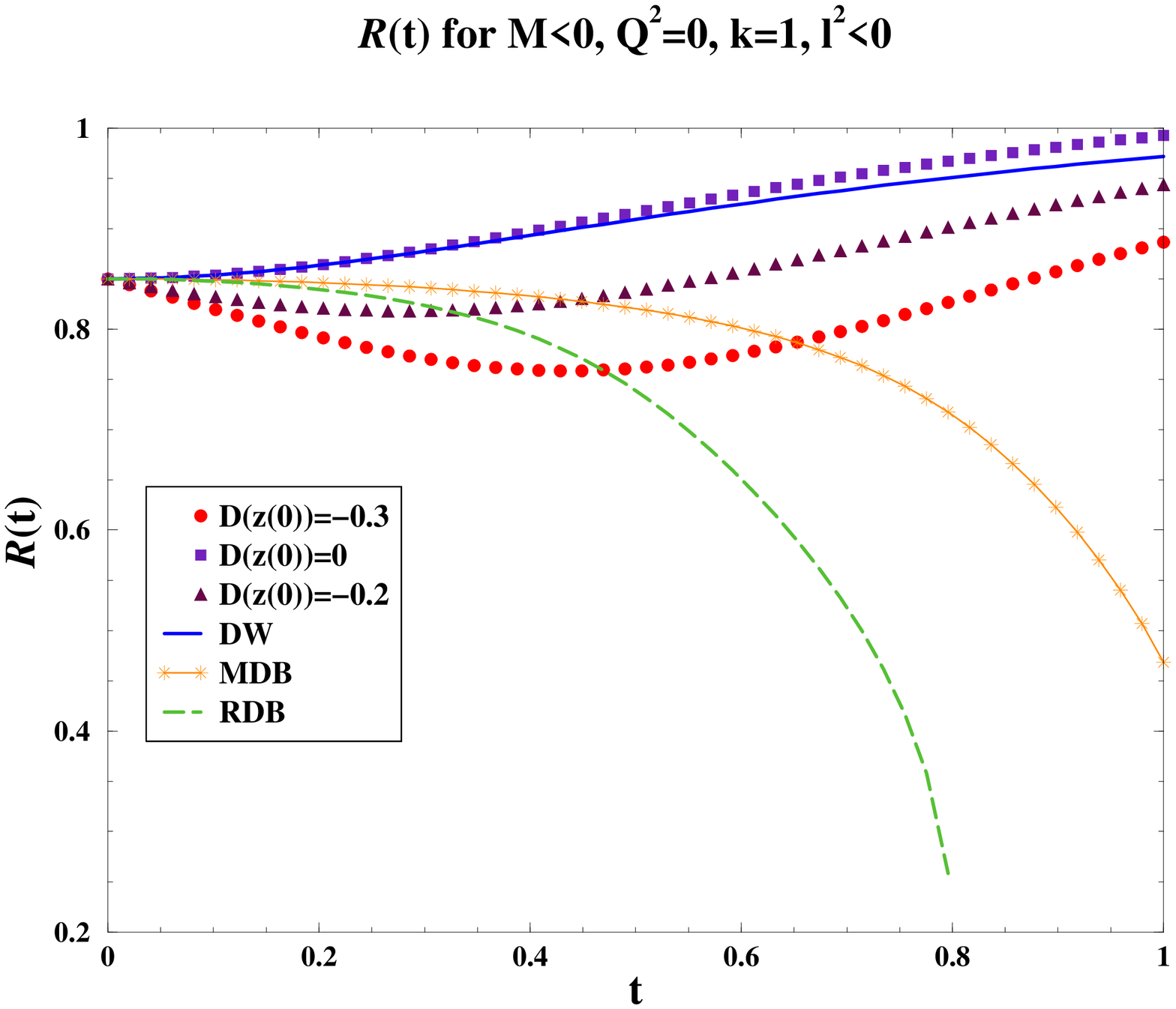}}\nonumber
\end{eqnarray}
\caption{Scale factor evolution for domain wall, matter and
radiation-dominated branes and geodesics when $M<0$, $Q^2=0$ and
$l^2<0$.}  
\label{sf4}
\end{center}
\end{figure*}

\begin{table}[htb!]
\begin{tabular}{|c|c|c|c|c|c|c|c|c|}\hline\hline
$M$ &$Q^2$&$k$ &$l^2$ &Bulk&DW&MDB&RDB&Geodesic\\ \hline\hline  
 +& +& {\footnotesize 0}& +&
{\footnotesize AdS-naked singularity}& {\footnotesize $\rightarrow
 \infty$/{\it B}}& {\footnotesize {\it B}/$\rightarrow 0$}& {\footnotesize
{\it B}/$\rightarrow 0$}&{\footnotesize {\it B}$^{*\dagger}$/$\rightarrow\infty
 ^\dagger$/$0$}\\
 +  & +  &{\footnotesize 1}   & +   &{\footnotesize AdS-naked
 singularity}&{\footnotesize $\rightarrow \infty$}&{\footnotesize
 $\rightarrow 0$}&{\footnotesize $\rightarrow 0$}&{\footnotesize {\it
 G}$^*$}\\  
 +  & +  &{\footnotesize -1}  & +   &{\footnotesize AdS-Top.ch.
 black hole}& {\footnotesize $\rightarrow r_H$}&{\footnotesize
 $\rightarrow r_H$}&{\footnotesize $\rightarrow
 r_H$}&{\footnotesize $\rightarrow r_H$} \\ 
 +  &$\ll$&{\footnotesize 0,-1}&+    &{\footnotesize AdS-Top.ch.
 black hole} & {\footnotesize $\rightarrow r_H$}&{\footnotesize
 $\rightarrow r_H$}&{\footnotesize $\rightarrow r_H$}
 &{\footnotesize $\rightarrow {r_H}^*$} \\ 
 +  &$\ll$&{\footnotesize 1}& +   &{\footnotesize
 AdS-Reissner-Nordstr\"om}&{\footnotesize $\rightarrow
 r_H$/$\infty$}&{\footnotesize $\rightarrow
 r_H$}&{\footnotesize $\rightarrow r_H$} &{\footnotesize $\rightarrow
 {r_H}^*$} \\   
 -  & +  &{\footnotesize 0,1} & +   &{\footnotesize AdS-naked
 singularity}&{\footnotesize $\rightarrow \infty$}&{\footnotesize
 $\rightarrow 0$}&{\footnotesize $\rightarrow 0$}&{\footnotesize {\it
 G}$^*$}\\  
 -  & +  &{\footnotesize -1}& +   &{\footnotesize AdS-Top.ch.
 black hole}&{\footnotesize $\rightarrow r_H$}&{\footnotesize
 $\rightarrow r_H$}&{\footnotesize $\rightarrow r_H$}
 &{\footnotesize $\rightarrow {r_H}^*$} \\ 
 +  & +  &{\footnotesize 0,-1}&- &{\footnotesize dS-naked
 singularity}&{\footnotesize $\rightarrow r_c$}&{\footnotesize
 $\rightarrow r_c$}&{\footnotesize $\rightarrow
 0$/$r_c$} &{\footnotesize $\rightarrow {r_c}^{*\dagger}$}\\ 
 +& +& {\footnotesize 1}& - &{\footnotesize dS-naked singularity}&
 {\footnotesize $\rightarrow r_c$}& {\footnotesize {\it
 B}/$\rightarrow 0$/$r_c$}&{\footnotesize $\rightarrow 
 0$/$r_c$} &{\footnotesize $\rightarrow {r_c}^{*\dagger}$}\\
+&$\ll$&{\footnotesize 0,-1}&-&{\footnotesize dS-naked
 singularity}&{\footnotesize $\rightarrow r_c$}&{\footnotesize
 $\rightarrow r_c$}&{\footnotesize $\rightarrow
 0$/$r_c$} &{\footnotesize $\rightarrow {r_c}^{*\dagger}$/$0$}\\ 
 +& $\ll$& {\footnotesize 1}& -&{\footnotesize
 dS-Reissner-Nordstr\"om}&{\footnotesize $\rightarrow
 r_H$/$r_c$}&{\footnotesize $\rightarrow 
 r_H$/$r_c$}&{\footnotesize $\rightarrow r_H$/$r_c$} &{\footnotesize
 $\rightarrow {r_H}^*$/${r_c}^{*\dagger}$}\\    
 -  & +  &{\footnotesize 0,$\pm$1}& -&{\footnotesize dS-naked
 singularity}&{\footnotesize $\rightarrow r_c$}&{\footnotesize $\rightarrow 
 0$/$r_c$}&{\footnotesize $\rightarrow 0$/$r_c$} &{\footnotesize
 $\rightarrow {r_c}^{*\dagger}$/$0$} \\ 
 \hline \hline
\end{tabular} 
\caption{Scale factor and geodesics evolution (charged case). {\it B} and
  {\it G} mean bouncing and growing behaviour, respectively. The $*$ or
  the $\dagger$  
in the last column indicates the possibility of shortcuts for the
matter and radiation-dominated branes or the domain wall, respectively.}
\label{sum2}
\end{table}

\begin{figure*}[htb!]
\begin{center}
\leavevmode
\begin{eqnarray}
\epsfxsize= 6.0truecm\rotatebox{-90}
{\epsfbox{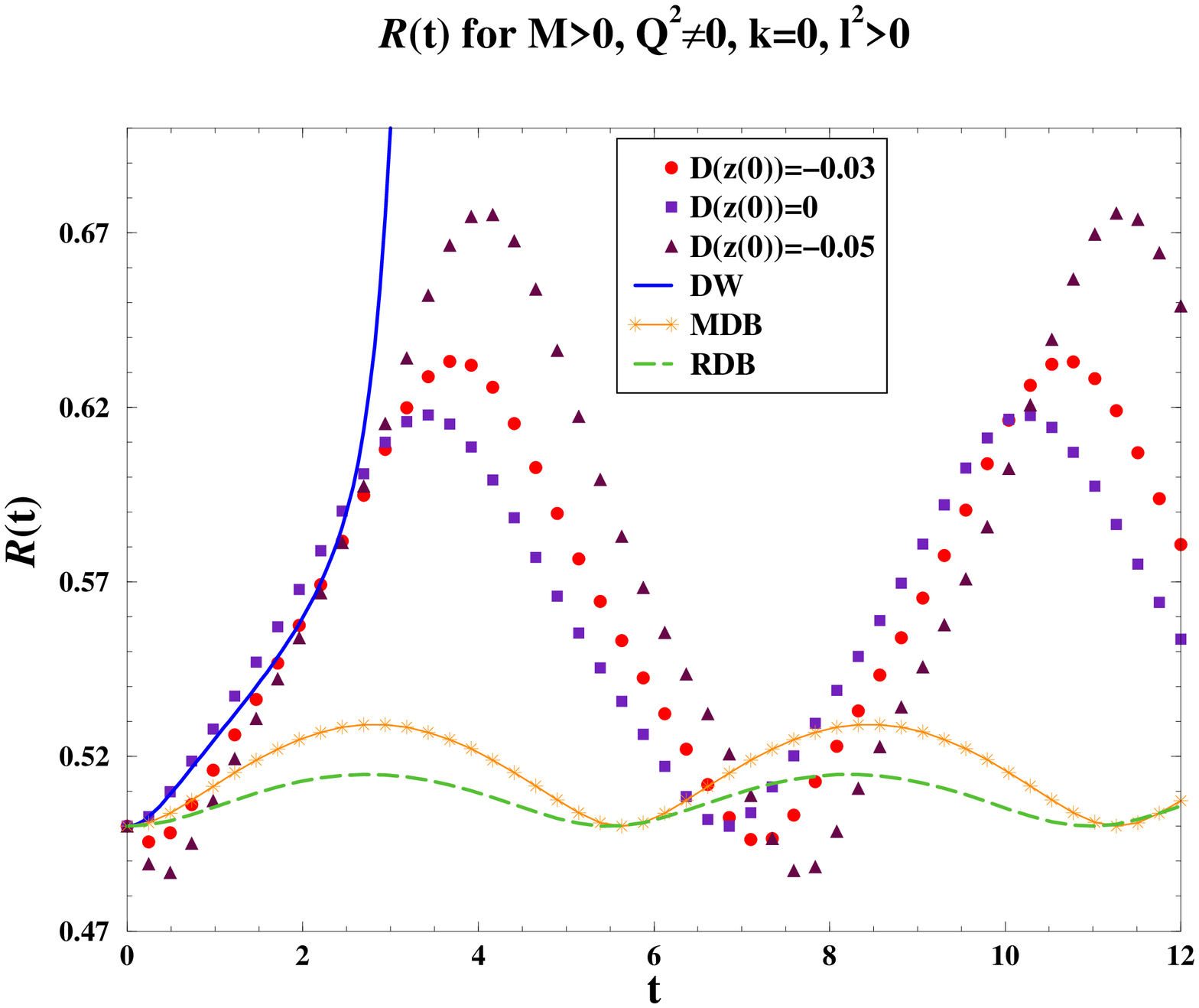}}\nonumber
\epsfxsize= 6.0truecm\rotatebox{-90}
{\epsfbox{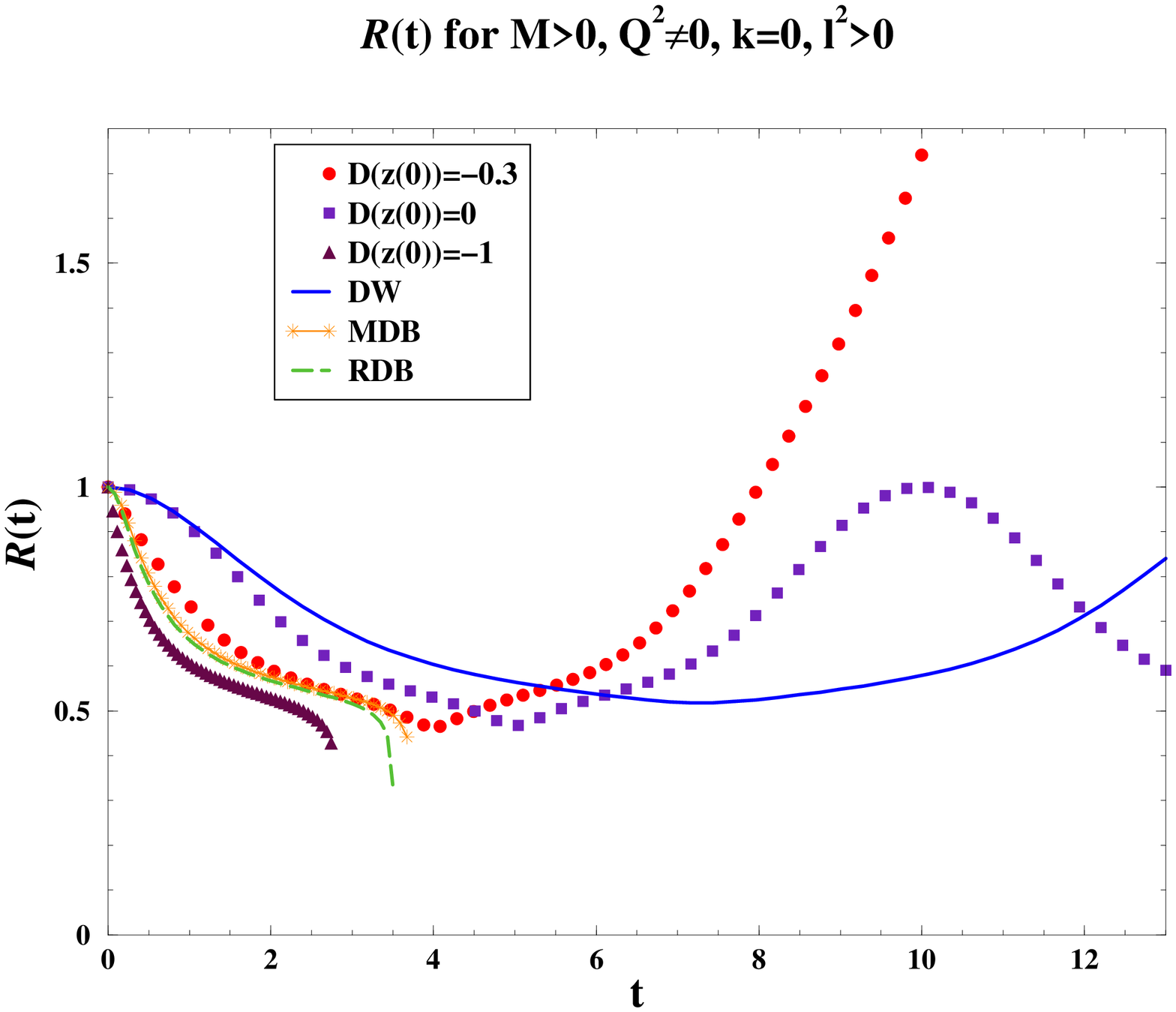}}\nonumber
\end{eqnarray}
\caption{Scale factor evolution for domain wall, matter and
radiation-dominated branes and geodesics when $M>0$, $Q^2 \not= 0$ and
$l^2>0$.} 
\label{sf5}
\end{center}
\end{figure*}
\begin{figure*}[htb!]
\begin{center}
\leavevmode
\begin{eqnarray}
\epsfxsize= 6.0truecm\rotatebox{-90}
{\epsfbox{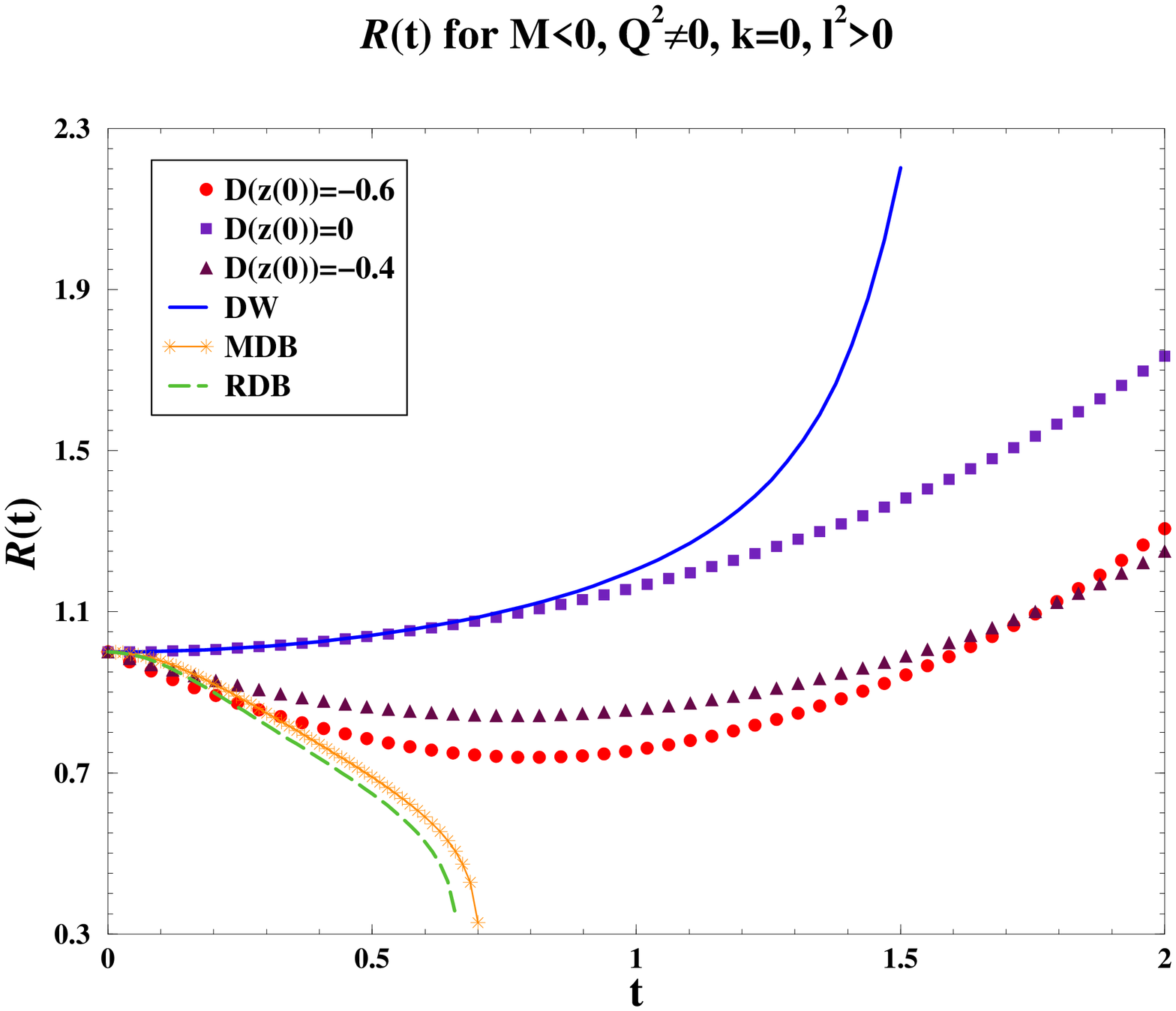}}\nonumber
\epsfxsize= 6.0truecm\rotatebox{-90}
{\epsfbox{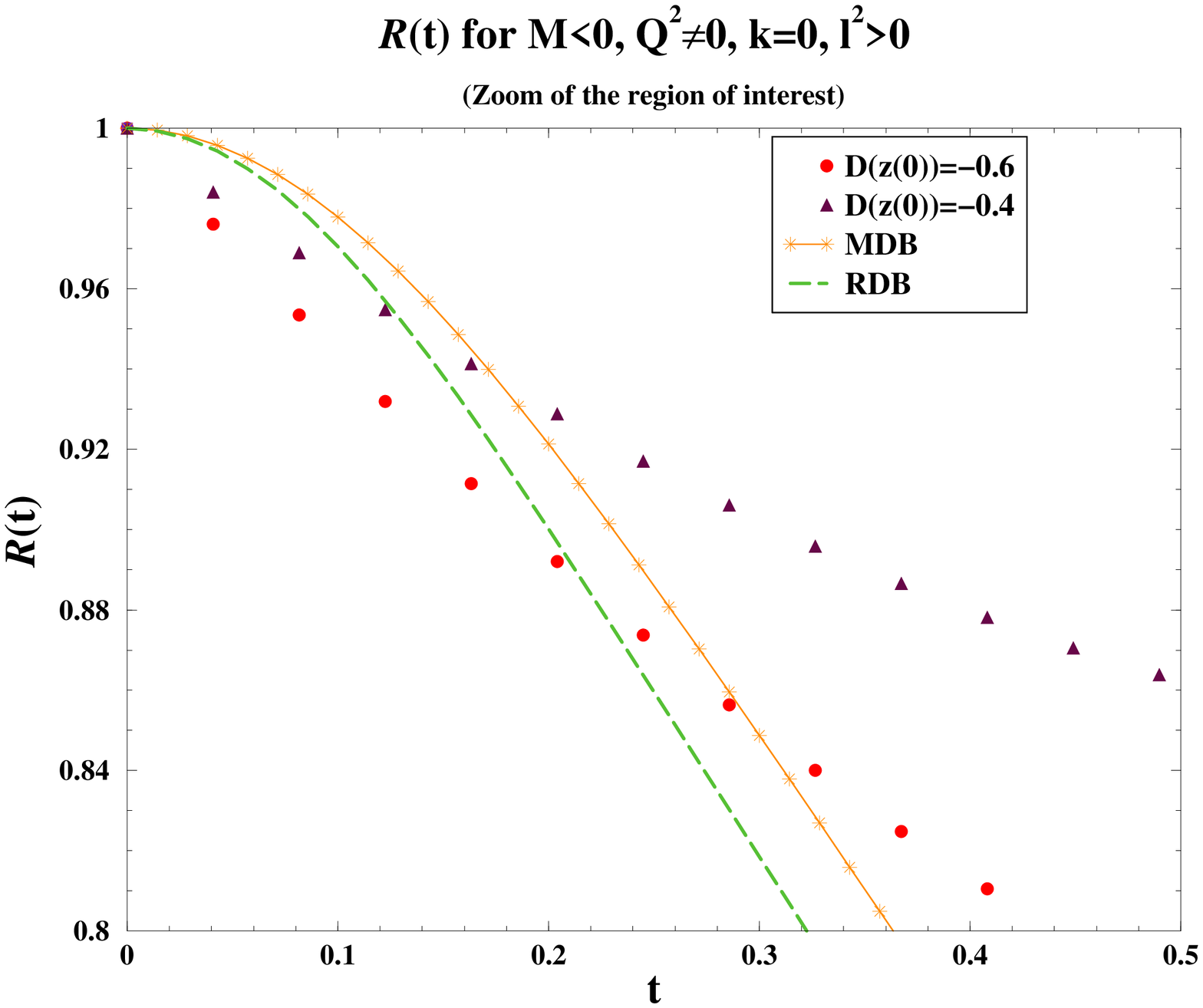}}\nonumber
\end{eqnarray}
\caption{Scale factor evolution for domain wall, matter and
radiation-dominated branes and geodesics when $M<0$, $Q^2 \not= 0$ and
$l^2>0$.} 
\label{sf6}
\end{center}
\end{figure*}
\begin{figure*}[htb!]
\begin{center}
\leavevmode
\begin{eqnarray}
\epsfxsize= 6.0truecm\rotatebox{-90}
{\epsfbox{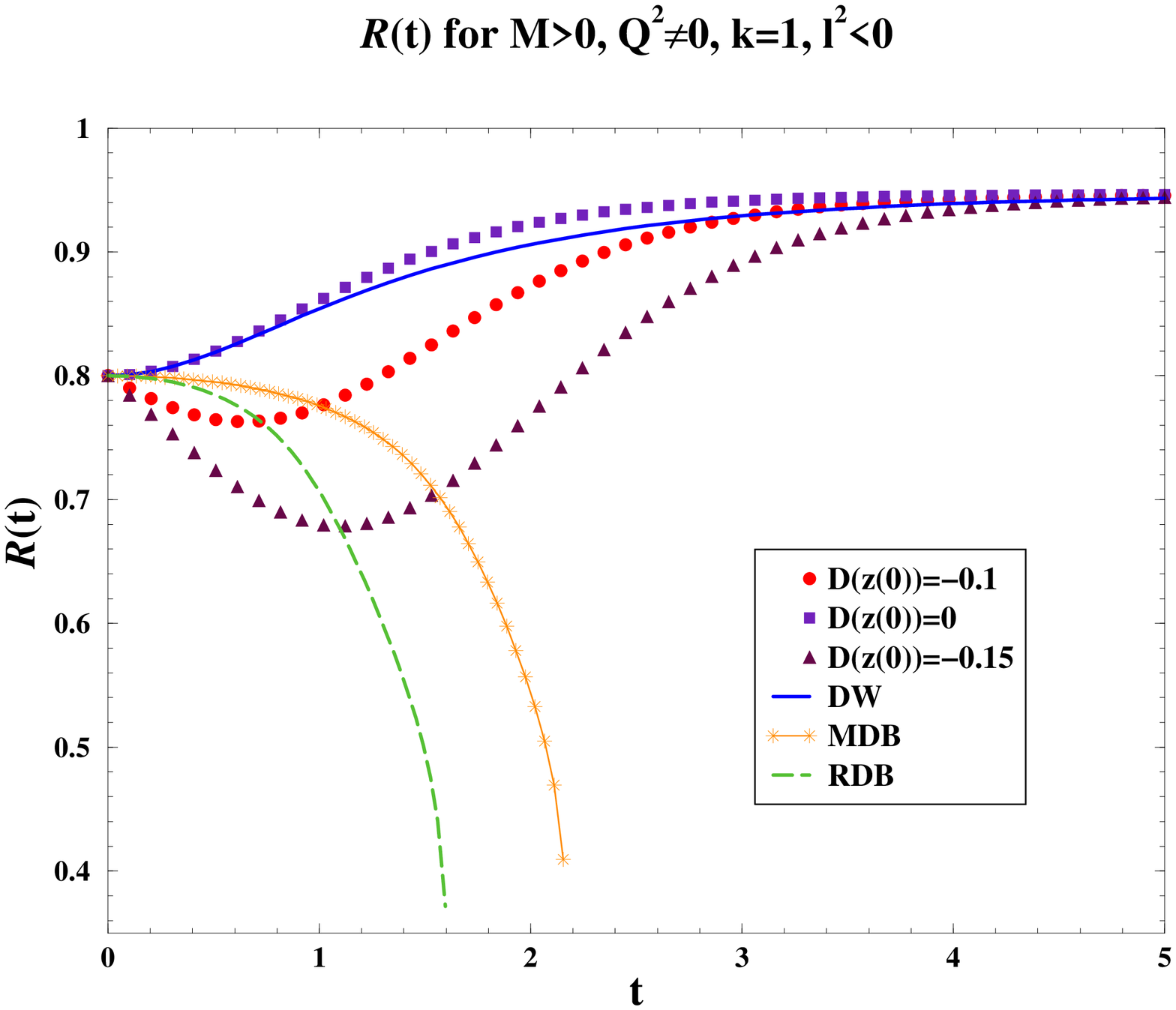}}\nonumber
\epsfxsize= 6.0truecm\rotatebox{-90}
{\epsfbox{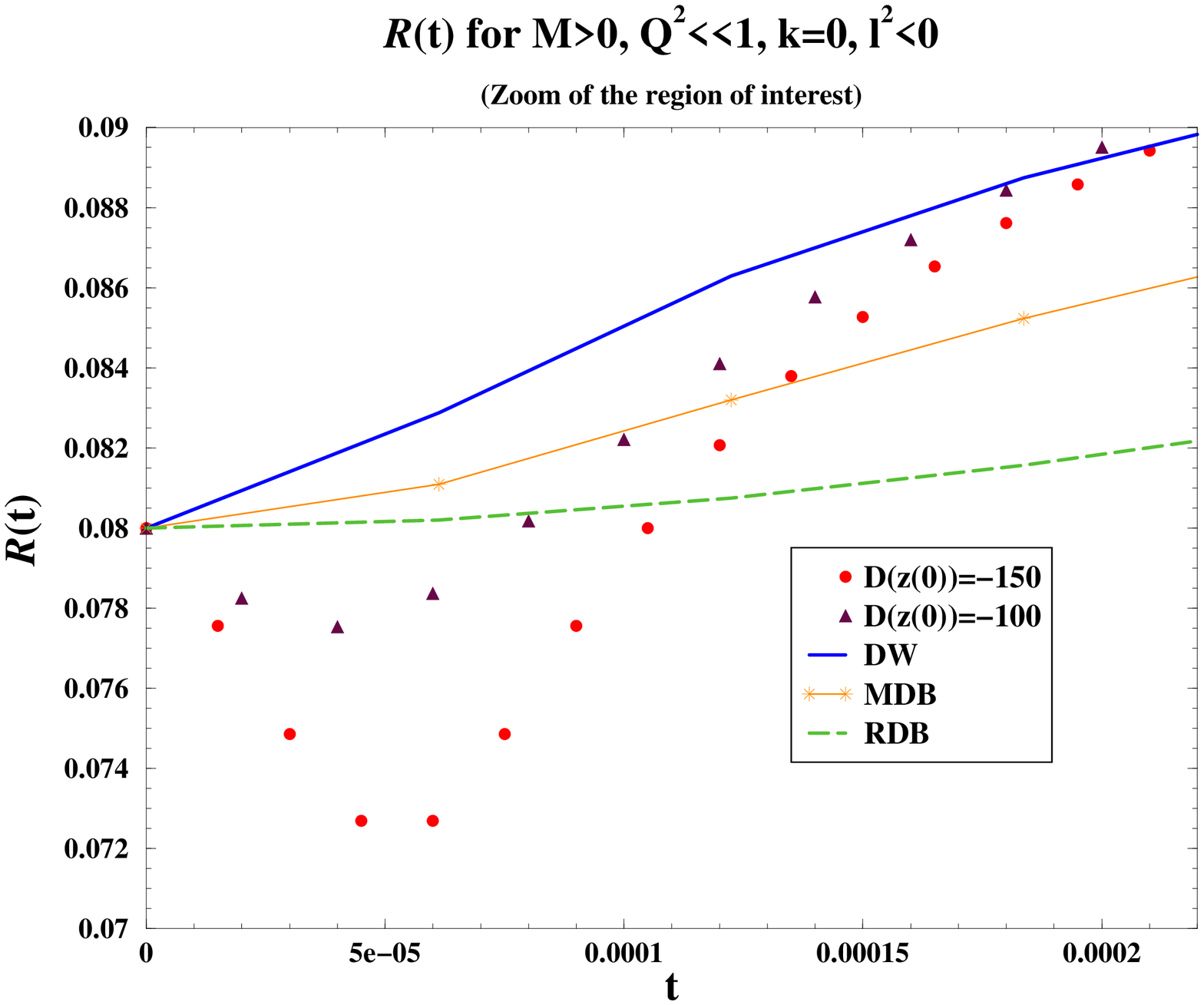}}\nonumber
\end{eqnarray}
\caption{Scale factor evolution for domain wall, matter and
radiation-dominated branes and geodesics when $M>0$, $Q^2 \not= 0$ and
$l^2<0$.}  
\label{sf7}
\end{center}
\end{figure*}
\begin{figure*}[htb!]
\begin{center}
\leavevmode
\begin{eqnarray}
\epsfxsize= 6.0truecm\rotatebox{-90}
{\epsfbox{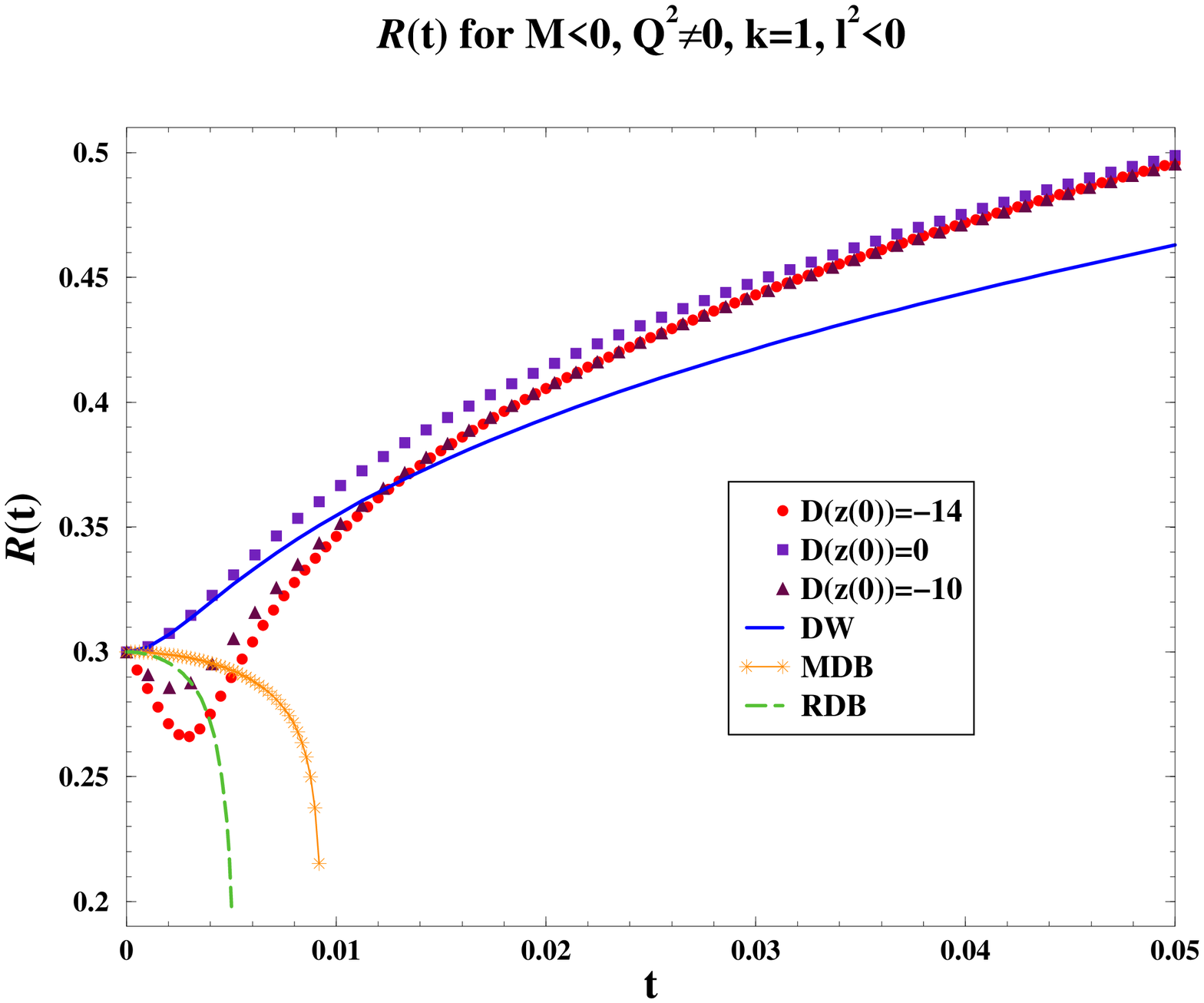}}\nonumber
\epsfxsize= 6.0truecm\rotatebox{-90}
{\epsfbox{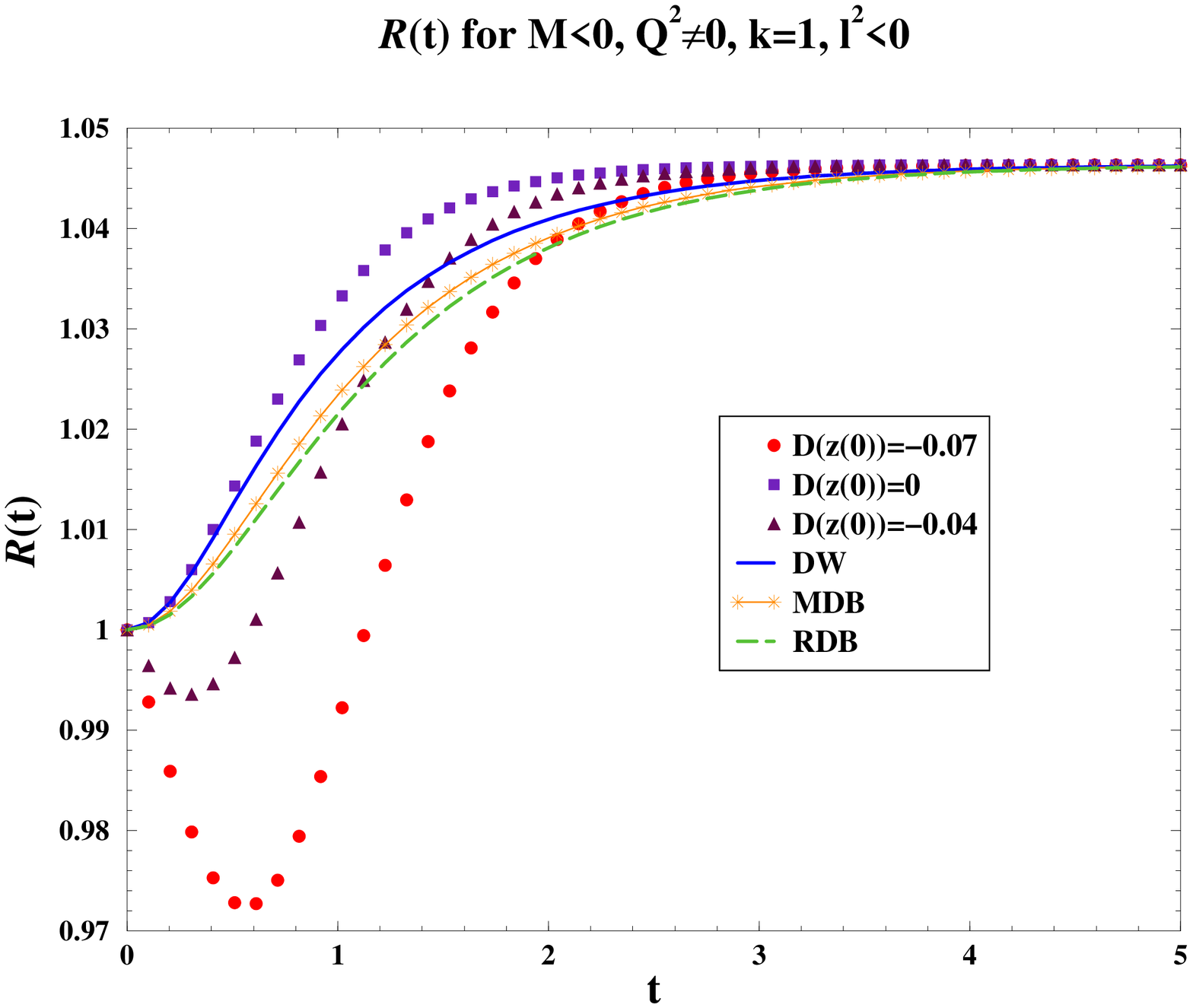}}\nonumber
\end{eqnarray}
\caption{Scale factor evolution for domain wall, matter and
radiation-dominated branes and geodesics when $M<0$, $Q^2 \not= 0$ and
$l^2<0$.}  
\label{sf8}
\end{center}
\end{figure*}
\begin{table}[t!]
\begin{tabular}{|c|c|c|c|c|c|c|c|c|c|c|c|c|}\hline\hline
&\multicolumn{4}{c|}{\bf DW} & \multicolumn{4}{c|}{\bf MDB} &
\multicolumn{4}{c|}{\bf RDB}\\ \hline
{\scriptsize Fig.}&$t$&$\tau$&$\Delta\tau$&$g/\gamma$&$t$&$\tau$&$\Delta\tau$&$g/\gamma$&$t$&$\tau$&$\Delta\tau$&$g/\gamma$ \\\hline\hline
{\footnotesize 2}&{\footnotesize -}     &{\footnotesize -}
&{\footnotesize -}    &{\footnotesize -}&{\footnotesize
.54}&{\footnotesize .408} &{\footnotesize .004} &{\footnotesize 1.011}
&{\footnotesize .27}&{\footnotesize .238}&{\footnotesize .002}
&{\footnotesize 1.009}\\ 
{\footnotesize 3}&{\footnotesize - }&{\footnotesize -}&{\footnotesize
-}    &{\footnotesize -} &{\footnotesize .23}   &{\footnotesize .25}
&{\footnotesize .01}  &{\footnotesize 1.047} &{\footnotesize
.18}&{\footnotesize .191}  &{\footnotesize .005} &{\footnotesize 1.033} \\
 
{\footnotesize 4}&{\footnotesize 2.39}  &{\footnotesize .563}
&{\footnotesize .009} &{\footnotesize 1.015} &{\footnotesize 1.27}
&{\footnotesize .393} &{\footnotesize .003} &{\footnotesize 1.007}
&{\footnotesize .95}   &{\footnotesize .306}  &{\footnotesize .002}
&{\footnotesize 1.006} \\  
{\footnotesize 4}&{\footnotesize 3.51}  &{\footnotesize .76}
&{\footnotesize .03}  &{\footnotesize 1.035} &{\footnotesize 2.33}
&{\footnotesize .67}  &{\footnotesize .02}  &{\footnotesize 1.024}
&{\footnotesize 1.74}  &{\footnotesize .56}  &{\footnotesize .01}
&{\footnotesize 1.019}\\  
{\footnotesize 5}&{\footnotesize 1.58}  &{\footnotesize .53}
&{\footnotesize .02}  &{\footnotesize 1.028} &{\footnotesize .89}
&{\footnotesize .412} &{\footnotesize .007} &{\footnotesize 1.017}
&{\footnotesize .66}   &{\footnotesize .326}  &{\footnotesize .004}
&{\footnotesize 1.013}\\  
{\footnotesize 5}&{\footnotesize 2.20}  &{\footnotesize .69}
&{\footnotesize .04}  &{\footnotesize 1.061} &{\footnotesize 1.45}
&{\footnotesize .63}  &{\footnotesize .03}  &{\footnotesize 1.046}
&{\footnotesize 1.06}  &{\footnotesize .54}   &{\footnotesize .02}
&{\footnotesize 1.039}\\  
{\footnotesize 5}&{\footnotesize - }    &{\footnotesize -}
&{\footnotesize -}    &{\footnotesize -} &{\footnotesize .66}
&{\footnotesize .47}  &{\footnotesize .03}  &{\footnotesize 1.064}
&{\footnotesize .47}   &{\footnotesize .32}   &{\footnotesize .01}
&{\footnotesize 1.046}\\ 
{\footnotesize 5}&{\footnotesize -}&{\footnotesize -}&{\footnotesize -}
&{\footnotesize -}&{\footnotesize .44}&{\footnotesize .300}
&{\footnotesize .006}&{\footnotesize 1.021}&{\footnotesize .33}
&{\footnotesize .224}&{\footnotesize .003} &{\footnotesize 1.015}\\  
{\footnotesize 6}&{\footnotesize 2.03}  &{\footnotesize .464}
&{\footnotesize .006}&{\footnotesize 1.013} &{\footnotesize .80}
&{\footnotesize .232} &{\footnotesize .002} &{\footnotesize 1.008}
&{\footnotesize .64} &{\footnotesize .192}  &{\footnotesize .001}
&{\footnotesize 1.007} \\   
{\footnotesize 6}&{\footnotesize -}     &{\footnotesize -}
&{\footnotesize -}    &{\footnotesize -} &{\footnotesize 1.11}
&{\footnotesize .318} &{\footnotesize .007} &{\footnotesize 1.022}
&{\footnotesize .94}   &{\footnotesize .283}  &{\footnotesize .006}
&{\footnotesize 1.020} \\  
{\footnotesize 6}&{\footnotesize 5.5}   &{\footnotesize 1.6}
&{\footnotesize .2}   &{\footnotesize 1.206} &{\footnotesize -}
&{\footnotesize -}    &{\footnotesize -}    &{\footnotesize -}
&{\footnotesize -}&{\footnotesize -}    &{\footnotesize -}
&{\footnotesize -} \\  
{\footnotesize 6}&{\footnotesize 6.2}   &{\footnotesize 2.5}
&{\footnotesize .1}   &{\footnotesize 1.074} &{\footnotesize -}
&{\footnotesize -}    &{\footnotesize -}    &{\footnotesize -}
&{\footnotesize -}     &{\footnotesize -}     &{\footnotesize -}
&{\footnotesize -}\\  
{\footnotesize 7}&{\footnotesize -}     &{\footnotesize -}
&{\footnotesize -}    &{\footnotesize -} &{\footnotesize .30}
&{\footnotesize .259} &{\footnotesize .006} &{\footnotesize 1.024}
&{\footnotesize .22}   &{\footnotesize .191}  &{\footnotesize .003}
&{\footnotesize 1.018}\\  
{\footnotesize 7}&{\footnotesize -}     &{\footnotesize -}
&{\footnotesize -}    &{\footnotesize -} &{\footnotesize .175}
&{\footnotesize .169} &{\footnotesize .002} &{\footnotesize 1.011}
&{\footnotesize .13}   &{\footnotesize .126}  &{\footnotesize .001}
&{\footnotesize 1.008} \\  
{\footnotesize 8}&{\footnotesize 3.07}  &{\footnotesize .91}
&{\footnotesize .06}  &{\footnotesize 1.065} &{\footnotesize 1.01}
&{\footnotesize .47}  &{\footnotesize .01}  &{\footnotesize 1.026}
&{\footnotesize .73}   &{\footnotesize .337}  &{\footnotesize .006}
&{\footnotesize 1.019}\\  
{\footnotesize 8}&{\footnotesize -}    &{\footnotesize -}
&{\footnotesize -}    &{\footnotesize -} &{\footnotesize 1.56}
&{\footnotesize .76}  &{\footnotesize .07}  &{\footnotesize 1.094}
&{\footnotesize 1.1}   &{\footnotesize .51}   &{\footnotesize .03}
&{\footnotesize 1.061} \\  
{\footnotesize 8}&{\scriptsize 2.2}{\tiny E}{\scriptsize
-4}&{\scriptsize 2}{\tiny E}{\scriptsize -3}&{\scriptsize
2}{\tiny E}{\scriptsize -4}&{\footnotesize 1.084} 
&{\scriptsize 1.4}{\tiny E}{\scriptsize -4}&{\scriptsize
1.7}{\tiny E}{\scriptsize -3}&{\scriptsize 1}{\tiny E}{\scriptsize
-4}&{\footnotesize 1.072} &{\scriptsize 1.1}{\tiny E}{\scriptsize
-4}&{\scriptsize 1.6}{\tiny E}{\scriptsize
-3} &{\scriptsize 1}{\tiny E}{\scriptsize -4}&{\footnotesize 1.069}\\  
{\footnotesize 8}&{\scriptsize 1.9}{\tiny E}{\scriptsize -4}&{\scriptsize
1.8}{\tiny E}{\scriptsize -3}&{\scriptsize 7}{\tiny
E}{\scriptsize -5} &{\footnotesize 1.035} 
&{\scriptsize 1.1}{\tiny E}{\scriptsize -4}&{\scriptsize
1.4}{\tiny E}{\scriptsize -3}&{\scriptsize 3}{\tiny E}{\scriptsize -5}
&{\footnotesize 1.024}&{\scriptsize 8.2}{\tiny E}{\scriptsize
-5}&{\scriptsize 1.1}{\tiny E}{\scriptsize -3}&{\scriptsize
2}{\tiny E}{\scriptsize -5}&{\footnotesize 1.022}\\ 
{\footnotesize 9}&{\footnotesize .013}  &{\footnotesize .036}
&{\footnotesize .004} &{\footnotesize 1.106} &{\footnotesize .0052}
&{\footnotesize .023} &{\footnotesize .002} &{\footnotesize 1.077}
&{\footnotesize .004}  &{\footnotesize .019}  &{\footnotesize .001}
&{\footnotesize 1.062} \\  
{\footnotesize 9}&{\footnotesize .012}  &{\footnotesize .035}
&{\footnotesize .002} &{\footnotesize 1.056}&{\footnotesize .0042}
&{\footnotesize .0186}&{\scriptsize 5}{\tiny E}{\scriptsize
-3}&{\footnotesize 1.030} &{\footnotesize .0031} &{\footnotesize .0139}
&{\scriptsize 3}{\tiny E}{\scriptsize -3}&{\footnotesize 1.023} \\  
{\footnotesize 9}&{\footnotesize 2.4}   &{\footnotesize .52}
&{\footnotesize .02}  &{\footnotesize 1.029} &{\footnotesize 2.2}
&{\footnotesize .52}  &{\footnotesize .01}  &{\footnotesize
1.027}&{\footnotesize 2.0}  &{\footnotesize .51}   &{\footnotesize
.01} &{\footnotesize 1.026}\\  
{\footnotesize 9}&{\footnotesize 1.5}   &{\footnotesize .355}
&{\footnotesize .004} &{\footnotesize 1.010} &{\footnotesize 1.3}
&{\footnotesize .339} &{\footnotesize .003} &{\footnotesize 1.008}
&{\footnotesize 1.1}   &{\footnotesize .314}  &{\footnotesize .002}
&{\footnotesize 1.007} \\ \hline\hline  
\end{tabular} 
\caption{Bulk time $t$, brane time $\tau$, time delays $\Delta\tau$
and ratio between graviton and photon horizons $g/\gamma$ for shortcut
geodesics.}  
\label{td}
\end{table}

First, let us consider the zero charge case. The following results
have also been summarized in table \ref{sum1} for a better comprehension
of the several solutions.

\subsection{$M>0$, $Q^2=0$, $l^2>0$}

For positive both $M$ and $l^2$ when $k=1$, the bulk is anti de
Sitter-Schwarzschild solution, while for $k \not= 1$ we have a
``topological'' black hole with a flat or hyperbolic event horizon in
an asymptotically anti de Sitter space. This case is shown in
figure \ref{hh}(a). 

For a domain wall ($\omega=-1$) and $k=1$ the brane falls into
the event horizon $r_H$ when the 
initial condition for the brane position is near $r_H$, 
otherwise ${\cal R}(t)$ grows. On the other hand, for $k \not= 1$ the
brane always falls into $r_H$. 

In the case of a matter-dominated brane ($\omega=0$) the brane falls
into the event horizon for any $k$. This behaviour is also verified for
a radiation-dominated brane ($\omega=1/3$).

Let us see now the geodesic behaviour. When $k=1$ and the initial
condition on the brane is chosen near 
$r_H$, the geodesics fall into the event horizon. Some of them follow
the domain wall at the beginning and others leave the matter or
radiation-dominated branes returning after a short time (shortcut
geodesics) and leaving again the branes to fall into the
singularity. Conversely, if the initial condition is taken far from
the event horizon, the geodesics grow accompanying the domain wall
initially, though their growth is slower; however, they never return
to it after their decoupling.
When $k\not= 1$, independently of the initial condition the geodesics
behave in the same way as the previous case when the initial condition
is taken near $r_H$ and we again find some shortcuts for the matter
and radiation-dominated branes before all the geodesics fall into the
event horizon. Some results are shown in figure \ref{sf1}.

\subsection{$M<0$, $Q^2=0$, $l^2>0$}

As we can see from figure \ref{hh}(b), for negative $M$ and positive
$l^2$ when $k=-1$, $h$ describes a topological black hole in an 
asymptotically anti de Sitter space with hyperbolic event horizon. If
$k \not= -1$, there is a timelike naked singularity and 
the metric is asymptotically anti de Sitter.

The solutions of the brane equation of motion (\ref{brem}) display the
following features. In the case of a domain wall, 
when $k=-1$, the brane falls into the event horizon; while if
$k\not= -1$, the solution for ${\cal R}(t)$ grows. 

For ($\omega=0$) the matter-dominated brane falls into the naked
singularity when $k \not= 
-1$ and into the event horizon if $k=-1$. The radiation-dominated
brane displays the same behaviour.

On the other hand, the solutions of equation (\ref{geo}) show that when $k\not=
-1$, the geodesics grow slower than the domain wall. Some 
of them are shortcuts for matter or radiation-dominated branes since
they leave and return to them before the branes reach the naked
singularity. Furthermore, if the geodesic (negative) initial velocity
is big enough, it can fall into the naked singularity.
Besides, when $k=-1$, all the geodesics fall into the event
horizon. We can see some results in figure \ref{sf2}.

\subsection{$M>0$, $Q^2=0$, $l^2<0$}

In the case of positive $M$, negative $l^2$ and $k=1$, the metric
is de Sitter-Schwarzschild with event and 
cosmological horizons given by the zeros of $h({\cal R})$. If $k \not=
1$, there is a cosmological singularity at ${\cal R}=0$ in
the asymptotically de Sitter background. See figure \ref{hh}(c). 

In the case of a domain wall for $k=1$, the solutions for ${\cal
R}(t)$ converge either to the 
event or the cosmological horizon. For $k \not= 1$, there is no
solution since ${\cal R}$ turns to be a time coordinate. The same
behaviour is observed for matter and radiation-dominated branes.

Since the solutions to the brane equation of motion just appear when
$k=1$, the solutions of the geodesic equation only have physical
meaning in this case. In this way we found that the geodesics
reach either the event or the cosmological horizon. In the latter case
there are many shortcuts for the matter or radiation-dominated branes
as well as for the domain wall before all of them reach the
cosmological horizon. Some results are illustrated in figure \ref{sf3}.

\subsection{$M<0$, $Q^2=0$, $l^2<0$}

When $M$ and $l^2$ are both negative, the bulk metric displays a
timelike naked singularity in 
asymptotically de Sitter space. There is also a cosmological horizon
which geometry is determined by $k$ as we can see from figure \ref{hh}(d).

In the case of a domain wall for any $k$ the solution ${\cal
R}(t)$ converges to the cosmological horizon.

When ($\omega =0$), the matter-dominated brane either falls into the
naked singularity or converges to the cosmological horizon
independently of the value of $k$. This same result was found for a
radiation-dominated brane.

On the other hand, the geodesics either fall into the naked
singularity when the initial velocity is negative enough or converge
to the cosmological horizon for any $k$. In the latter case we found
several shortcuts for the domain wall and the branes. We can see some
results in figure \ref{sf4}. 

\bigskip

Now let us consider the charged solutions. As in the uncharged case,
we also show a summary of our results in table \ref{sum2}.

\subsection{$M>0$, $Q^2 \not= 0$, $l^2>0$}

When $M$ and $l^2$ are both positive, for $k=-1$ the metric describes a
topological charged black hole in 
asymptotically anti de Sitter space with hyperbolic horizon. While for $k
\not= -1$, there is a timelike naked singularity and the bulk is
asymptotically anti de Sitter (see figure \ref{hh}(e)). Note that if the
charge is small, all the metrics describe topological charged black
holes in anti de Sitter bulks. In particular, when $k=1$, we have anti de
Sitter-Reissner-Nordstr\"om bulk. 

Let us see the solutions of the brane equation of motion. In the case
of a domain wall there are different behaviours for each value of
$k$. When $k=0$, 
${\cal R}(t)$ bounces for the initial condition ${\cal R}(0)$ greater
than or equal to the minimum of $h({\cal R})$; otherwise, it diverges to 
infinity. When $k=1$, ${\cal R}(t)$ diverges to infinity. When $k=-1$,
the domain wall falls into the event horizon. 
For small charges ${\cal R}(t)$ converges to the event horizon except
when $k=1$, where it can also grow or diverge to infinity after a certain
initial value. 
In the limit $Q \rightarrow 0$ we also verified that our solutions
converge to those in section 4.1.

For a matter-dominated brane ($\omega=0$) we again found different
behaviours for each $k$. When $k=0$, ${\cal 
R}(t)$ bounces for the initial condition ${\cal R}(0)$ less than or equal
to the minimum of $h({\cal R})$; otherwise, it falls into the naked
singularity. When $k=1$, ${\cal R}(t)$ goes to the naked
singularity. When $k=-1$, it falls into the event horizon.
For small charges the brane always falls into the event
horizon. Again when $Q \rightarrow 0$ we recover the solutions in
section 4.1. 

In the case of a radiation-dominated brane ($\omega=1/3$) the behaviour
of the solution is somewhat similar to the matter-dominated brane case.
When $k=0$, ${\cal R}(t)$ bounces for the initial condition ${\cal
R}(0)$ less than or equal to the minimum of $h({\cal R})$, but it falls
into the naked singularity when the initial condition is near
$0$ or when it is greater than the minimum of $h({\cal R})$. When
$k=1$, ${\cal R}(t)$ goes to the naked singularity. When $k=-1$, it
falls into the event horizon as the previous case. 
For small charges we have the same behaviour as in the matter-dominated
brane case.

Now let us investigate the geodesic behaviour. When $k=0$, the
geodesics bounce for the initial condition less than or 
equal to the minimum of $h({\cal R})$ in the same way as matter or
radiation-dominated branes producing several shortcuts in all brane
cases. However, when the initial condition is greater than the minimum
of $h({\cal R})$, the geodesics can continue to bounce or escape after
one oscillation or even fall into the naked singularity as a result of
increasing the initial (negative) velocity; in this case we just found
shortcuts for the domain wall as long as the geodesics complete at
least one oscillation. For small charges an event horizon is formed
and the geodesics fall into it but there are still some small
shortcuts for matter or radiation-dominated branes. 
When $k=1$, the geodesics grow slower than the domain wall producing
shortcuts just for matter or radiation-dominated branes. For small
charges the geodesics either diverge or fall into the event horizon.
When $k=-1$, all the geodesics fall into the event horizon as the
branes and there are no shortcuts. For small charges the geodesics
fall into the event horizon but can yield some shortcuts in the very
beginning of their paths for radiation or matter-dominated branes.
Some of these cases appear in figure \ref{sf5}.

\subsection{$M<0$, $Q^2 \not= 0$, $l^2>0$}

For negative $M$, positive $l^2$ and $k=-1$, the metric describes a
topological charged black hole in 
asymptotically anti de Sitter spacetime with hyperbolic horizon. While for
$k \not= -1$ there is a timelike naked singularity and the bulk
is asymptotically anti de Sitter as is shown in figure \ref{hh}(f). 

In the case of $\omega=-1$ and for $k=-1$ the domain wall falls into
the event horizon. If in turn $k \not= -1$, the solution ${\cal R}(t)$
diverges to infinity.  

When we consider a matter-dominated brane for $k=-1$, the brane
behaves in the same way as a domain wall; 
however, if $k \not= -1$, it falls into the naked singularity. This
same behaviour is observed for a radiation-dominated brane.

As for the solutions to the geodesic equation, when $k=-1$, the
geodesics fall into the event horizon and some of 
them can produce shortcuts for radiation or matter-dominated branes
before their falling.
If $k\not= -1$, the geodesics grow and yield shortcuts for matter or
radiation-dominated branes. Some results are shown in figure \ref{sf6}.

\subsection{$M>0$, $Q^2 \not= 0$, $l^2<0$}

In the case of positive $M$ and negative $l^2$ the metric describes a
timelike naked singularity in 
asymptotically de Sitter space with a cosmological horizon as we see
in figure \ref{hh}(g). However,
for a small charge the $k=1$ metric turns out to be de
Sitter-Reissner-Nordstr\"om with Cauchy, event and cosmological
horizons as it is shown in figure \ref{hh}(h).

For a domain wall the solution ${\cal R}(t)$ converges to the
cosmological horizon for any 
$k$. For a small charge this behaviour also applies except for $k=1$,
when the domain wall can also fall into the event horizon.

The case of a matter-dominated brane displays an interesting behaviour.
When $k=1$, ${\cal R}(t)$ bounces when the initial condition ${\cal
R}(0)$ is less than or equal to the only saddle point in $h({\cal
R})$; otherwise, it falls into the naked singularity or converges to
the cosmological horizon if the initial condition ${\cal R}(0)$ is
very near it. When $k \not=1$, the solutions converge to the
cosmological horizon.
For a small charge the solutions converge to the cosmological horizon
and also to the event horizon in the case $k=1$.

When we consider a radiation-dominated brane, we see that it either
falls into the naked singularity or 
converges to the cosmological horizon for any $k$.
For a small charge the solutions behave in the same way except if
$k=1$ when the brane converges either to the cosmological or the event
horizon.

Let us see the geodesic behaviour. For any $k$ all the geodesics
converge to the cosmological horizon and 
produce several shortcuts for all brane cases. 
When the charge is very small, an event horizon appears in the case
$k=1$ and the geodesics can either converge to the cosmological or the
event horizons. The geodesics converging to the cosmological horizon
produce shortcuts for all brane cases, while the geodesics falling
into the event horizon yield shortcuts just for matter or
radiation-dominated branes. On the other hand, when $k\not=1$, the
geodesics 
converge to the cosmological horizon after producing some shortcuts
for all brane cases, unless their initial (negative) velocity reaches
a threshold after which they fall into the naked singularity. Some
results are illustrated in figure \ref{sf7}.

\subsection{$M<0$, $Q^2 \not= 0$, $l^2<0$}

When $M$ and $l^2$ are both negative, the metric describes a timelike
naked singularity in asymptotically 
de Sitter space for any $k$ (see figure \ref{hh}(i)).

As for the solutions of the brane equation of motion, in the case of a
domain wall all the solutions converge to the cosmological horizon.

When we consider a matter-dominated brane, ${\cal R}(t)$ either falls
into the naked singularity or 
converges to the cosmological horizon for any $k$. The same behaviour
applies for a radiation-dominated brane.

As for the geodesics, they converge to the cosmological horizon and
produce several shortcuts for all brane cases for any $k$. However,
when the initial condition is taken near the singularity and the
initial velocity is negative enough, the geodesics can reach the
singularity and no shortcut appears. We can see some results in
figure \ref{sf8}.

\section{Discussion and Conclusions}

In the present work we have studied the behaviour of a brane embedded
in a six-dimensional de Sitter or anti de Sitter spacetime containing
a singularity covered by at least one horizon in the case of black
hole type solutions or a timelike naked singularity. The system
of equations describing this behaviour from the point of view of an
observer in the bulk appears to be highly nonlinear. Before
numerically solving this system we have considered fluctuations about a
fixed brane position in order to have a better insight of the whole
problem. We have concluded that the case of a flat domain wall with
vanishing effective cosmological constant reproduces the equation of
motion for a free scalar field as in five dimensions \cite{radion,cgr}. The
``equilibrium'' position can be chosen arbitrarily but there is no
stability. 

By solving the full nonlinear system we found different behaviours for
the several scenarios appearing due to all the combinations of $M$,
$Q^2$, $k$ and $l^2$ taken into account. We chose some typical
values for $\omega$, i.e. domain wall, matter and radiation-dominated
branes, in order to illustrate the solutions. The results show branes
getting away from the singularity, falling into it, converging to
cosmological horizons when they exist or even bouncing between a
minimum and maximum values.

The bouncing behaviour found in some of our solutions of the brane
equation of motion appears to be in good agreement with the recent
investigations in five dimensions \cite{chargedbh},
where universes bouncing from a contracting to an expanding phase
without encountering past and/or future singularities appear. In this
way these results could provide support for a singularity-free
cosmology or to the so-called cyclic universe scenarios \cite{st}.

Finally, we also studied the geodesic behaviour in every scenario found
in the present work. Contrarily to the case of a static brane, where
shortcuts appeared under very restrictive conditions \cite{accm}, the
present model of a dynamic brane embedded in a static bulk displays
shortcuts in almost all cases and under very mild
conditions. Moreover, despite the fact that the time delay between
graviton and photon flight time is not percentually so big as in other
models \cite{acm} (what is also evident from the ratio between
graviton and photon horizons), it exists and can eventually be measured by the
brane observer, although further considerations are certainly needed in a 
stricter realistic model. On the other hand, the fact that shortcuts are
abundant in the studied setups lends further support to the idea of
solving the horizon problem via thermalization by graviton exchange
\cite{acm,ac}; however, we should stress that this is not a proof of
the solution of the problem yet.

\bigskip
{\bf Acknowledgements:} I would like to thank Elcio Abdalla for useful
discussions and for reading the manuscript. This work has been
supported by Funda\c c\~ao de Amparo \`a Pesquisa do Estado de S\~ao
Paulo {\bf (FAPESP)}, Brazil.

\end{document}